\documentclass[referee]{aa}

\usepackage{natbib}
\usepackage{aas_macros}
\usepackage[breaklinks=true,a4paper=true,pagebackref=true]{hyperref}

\usepackage{graphicx}
\usepackage[T1]{fontenc}
\usepackage{amsmath}
\usepackage{amssymb}
\usepackage{amsmath}
\usepackage{graphicx}
\usepackage{amsmath}
\usepackage{amssymb}
\usepackage{amsfonts}
\usepackage{graphicx,epsfig,graphics}
\usepackage{aas_macros}
\usepackage{graphics}
\usepackage{color}
\usepackage{epstopdf}
\usepackage{epsfig,psfig,ulem}

\def\be{\begin{eqnarray}}
\def\ee{\end{eqnarray}}
\graphicspath{{}}

\newcommand{\cjb}[1]{#1}

\graphicspath{./Figures/} 

\titlerunning{Polarized CMB recovery with sparse component separation}
\authorrunning{Bobin et al}

\title{Polarized CMB map recovery with sparse component separation}
\author{ \hspace{0.25in} J. Bobin\inst{1}, F. Sureau\inst{1}, J.-L. Starck\inst{1}}
\institute{$^1$ Laboratoire CosmoStat, AIM, UMR CEA-CNRS-Paris 7, Irfu, SAp/SEDI, Service d'Astrophysique, CEA Saclay, F-91191 GIF-SUR-YVETTE CEDEX, France.}

\begin{document}
 

\abstract{The polarization modes of the cosmological microwave background are an invaluable source of information for cosmology, and a unique window to probe the energy scale of inflation. Extracting such information from microwave surveys requires disentangling between foreground emissions and the cosmological signal, which boils down to solving a component separation problem. Component separation techniques have been widely studied for the recovery of CMB temperature anisotropies but quite rarely for the polarization modes. In this case, most component separation techniques make use of second-order statistics to discriminate between the various components. More recent methods, which rather emphasize on the sparsity of the components in the wavelet domain, have been shown to provide low-foreground, full-sky estimate of the CMB temperature anisotropies. Building on sparsity, the present paper introduces a new component separation technique dubbed PolGMCA (Polarized Generalized Morphological Component Analysis), which refines previous work to specifically tackle the estimation of the polarized CMB maps: i) it benefits from a recently introduced sparsity-based mechanism to cope with partially correlated components, ii) it builds upon estimator aggregation techniques to further yield a better noise contamination/non-Gaussian foreground residual trade-off. The PolGMCA algorithm is evaluated on simulations of full-sky polarized microwave sky simulations using the Planck Sky Model (PSM), which show that the proposed method achieve a precise recovery of the CMB map in polarization with low noise/foreground contamination residuals. \cjb{It provides improvements with respect to standard methods, especially on the galactic center where estimating the CMB is challenging.}}

\date{Received -; accepted -}
\maketitle

\section{Introduction}
\label{sec:intro}

The CMB provides a snapshot of the state of the Universe at the time of recombination. It therefore carries invaluable information about the infancy of the Universe and its evolution to the current state. After a series of full-sky measurements of the microwave sky (COBE \citep{1996ApJ...464L...1B}, WMAP \citep{WMAP9_1}), Planck \citep{PR1_compsep} has recently released an accurate estimation of the CMB temperature map. Studying the statistical properties of the CMB temperature maps already provides a huge amount of information about the primordial state of the Universe. Additionally, the CMB map is a radiation with linear polarization that is described by the three polarization modes T, E and B \citep{Hu_PolaPrimer}; these three modes are observed through the Stokes parameters T, Q and U. The information carried by CMB polarization is crucial for: i) improving the estimation of the cosmological parameters and break some of their degeneracies, ii) give a privileged access to probe the energy scale of inflation through the estimation of the scalar-to-tensor ratio $r$ and the tensor spectral index $n_t$. Indeed, inflationary models predict the presence of tensor perturbations, which would be traced by the presence of specific CMB B-modes.\\
From the first full-sky observations of CMB temperature anisotropies by Cosmic Background Explorer (COBE - see \citep{1996ApJ...464L...1B}) to the Planck 2013 results \citep{PR1_compsep}, a lot of attention has been paid to the estimation of temperature anisotropies by component separation and little to the polarization signal. In contrast to temperature, polarization anisotropies are more challenging to measure as their level is about $5$ orders of magnitude lower than the temperature anisotropies level.\\
Polarization anisotropies have first been measured by balloon \citep{Boomerang2005} and ground-based experiments such as CBI \citep{CBI_Pola}. The WMAP experiments more recently provided full-sky observations of the polarization anisotropies \citep{WMAP9_1}. The BICEP2 collaboration initially announced the detection of inflationary B-modes and a measurement of the scale-to-tensor ratio $r  = 0.2$, which would be in slight tension with the Planck 2013 and 2015 results \citep{PR2_Params} that puts a stronger constraint $r < 0.11$ ($95 \%$ confidence interval). If this discrepancy could have found various cosmological explanations, it was soon advocated a polarized foregrounds origin (see \citep{Flauger14} and \cjb{\citep{Mortonson14}}). This point was later supported by a study of dust polarization made available by the Planck consortium \citep{PR1_DustPola} and joint BICEP2/Planck investigations \citep{BICEP2_Planck}. This controversy highlights the central role played by foreground cleaning and component separation to provide an accurate estimation of polarized CMB maps.\\
Component separation methods have mainly focused on temperature analysis lately \citep{Leach08,NILC12,2012arXiv1206.1773B,WMAP9_LGMCA,PR1_LGMCA}. Most component separation techniques rely on a linear mixture model. According to this model, each observation of the sky $x_i$ at wavelength $\nu_i$ is modeled as the linear combination of $n$ components $\{s_j\}_{j=1,\cdots,n}$ :
\begin{equation}
\label{eq:mixtmodel}
\forall i=1,\cdots,M; \quad x_i = \sum_{j=1,\cdots,n} a_{ij} s_j + n_i,
\end{equation}
where $a_{ij}$ stands for the contribution of component $s_j$ in observation $x_i$ and $n_i$ models instrumental noise. In this model, the data are assumed to be at the same resolution. This model can be more conveniently recast in matrix form:
\begin{equation}
\label{eq:mixtmodel2}
{\bf X} = {\bf AS} + {\bf N}.
\end{equation}
where each observation $x_i$ is stored in the $i$-th row of the observation matrix $\bf X$, $\bf S$ is the source matrix, $\bf A$ is the mixing matrix and $\bf N$ models instrumental noise. From a physical viewpoint, the $j$-th column of $\bf A$, which we denote by $a^j$, contains the electromagnetic spectrum of the source $s_j$. \\
Better known as blind source separation (BSS) in statistics and signal processing \citep{JutBook}, component separation aims at estimating the mixing  matrix $\bf A$ and the sources $\bf S$ from the knowledge of the observations $\bf X$. This is a classical ill-posed inverse problem, which admits an infinite number of solutions. Disentangling physically relevant components from erroneous estimates further requires imposing some desired properties on the components to be retrieved. In the framework of BSS, component separation methods generally enforce some diversity or contrast to distinguish between distinct components. To the best of our knowledge, all standard component separation methods for polarized CMB observations build upon second-order statistics \citep{ILC_eriksen,WMAP9_1,WMAP_Naselsky_Pola,PolEMICA}.\\
Lately, building on different grounds, a novel component separation technique, LGMCA (Local-Generalized Morphological Component Analysis - see \citep{2012arXiv1206.1773B}) has been introduced. In contrast to standard component separation techniques, the LGMCA algorithm relies on the sparse representation of galactic foregrounds, which makes it more sensitive to the higher-order statistics of these components. Using GMCA, a joint WMAP-Planck high quality full-sky CMB map has been reconstructed. Additionally, this map has been shown to be free of detectable thermal SZ (Sunyaev-Zel'Dovich emission \cjb{\citep{SZoriginal}}), and with low foreground contamination \citep{PR1_LGMCA}. The resulting CMB temperature map has been used to analyze the large-scale anomalies of the CMB radiation. Thanks to its low level of foregrounds, masking of the galactic center was not necessary for CMB large scale studies \citep{PR1_LGMCA_anomalies,BenDavid14,Aiola14,Lanusse14}.

\subsection*{Contributions}
In this paper, building upon the concept of sparse signal representation, we introduce an extension of the LGMCA algorithm to compute an accurate estimate of the CMB polarized maps with improved noise/non-Gaussian foreground contamination trade-off. In contrast to the processing of CMB temperature data, the estimation of the CMB polarization anisotropies is an even more challenging task since the polarization signal is weak and dominated by galactic foregrounds and noise contamination. For that purpose, the proposed method, dubbed PolGMCA (Polarized GMCA), builds upon two new key mechanisms: i) it makes profit of recent advances in blind source separation (BSS) that allow for the efficient separation of sparse and partially correlated sources, ii) it builds upon map estimator aggregation, which helps providing a better balance between foreground residuals and noise contamination. The PolGMCA algorithm is detailed in Section \ref{sec:compsep}. Numerical experiments based on the Planck Sky Model are described in Section~\ref{sec:results}, which demonstrate the ability of the proposed method to achieve good separation performances on the galactic center as well as at large scales.

\section{Polarized Sparse Component Separation}
\label{sec:compsep}

\subsection{Component separation for polarized data}

\paragraph{Modeling of polarized data:}

Describing polarized fields can be done in several ways. They are generally measured by the Stokes parameters Q and U, which can further be expanded in the Harmonic domain as follows:
\begin{equation}
Q \pm iU = {\sum_{\ell,m}} \, _{\pm 2} a_{\ell m} \, _{\pm 2}Y_{\ell m},
\end{equation}
where $_{\pm 2}Y_{\ell m}$ stands for the spin-2 spherical harmonics basis functions.\\
Extending the concept of the wavelet representation for polarization vector fields on the sphere has been proposed in \citep{starck:pola09} based on the celebrated "a trous" algorithm \citep{Starck05}. Building a wavelet transform for Q/U polarization fields then consist in building $J$ successive smooth approximations of the Q and U maps via a convolution with dyadically rescaled versions of the so-called scaling function $h$:
\begin{eqnarray}
\forall j=0,\cdots,J ; & \; & c_j^Q = h_j \star Q \\
& & c_j^U = h_j \star U,
\end{eqnarray}
where $J$ stands for the total number of scales. Following \cite{Starck05}, the scaling function is built so as to verify isotropy in the pixel domain. The wavelet coefficients at scale $j$ are then defined as the difference of two consecutive approximations:
\begin{eqnarray}
\forall j=0,\cdots,J; & \;& \quad w_{j+1}^Q = c_j^Q - c_{j+1}^Q \\
 & & \quad w_{j+1}^U = c_j^U - c_{j+1}^U 
\end{eqnarray}
In practice, these operations boil down to simple multiplications of the Q and U maps spherical harmonics coefficients with the scaling filters $\{ h_j \}_{j=0,\cdots,J}$.\\
According to the \textit{\`a trous} algorithm, the Q and U maps can be reconstructed by a very simple summation:
\begin{eqnarray}
Q & = & c_J^Q + \sum_{j = 1}^{J} w_j^Q \\
U & = & c_J^U + \sum_{j = 1}^{J} w_j^U
\end{eqnarray}
The same polarization field can equivalently be described as E and B polarization modes \citep{zalda}:
\begin{eqnarray}
\label{eq:EBfield}
E = \sum_{\ell,m} a_{\ell m}^E Y_{\ell m} = \sum_{\ell,m} -\frac{1}{2} \left( {}_2 a_{\ell m} + {}_{-2} a_{\ell m} \right) \\
B = \sum_{\ell,m} a_{\ell m}^B Y_{\ell m} = \sum_{\ell,m} \frac{1}{2} i \left( {}_2 a_{\ell m} - {}_{-2} a_{\ell m} \right),
\end{eqnarray}
where $Y_{\ell m}$ stands for the standard 0-spin spherical harmonics functions. The E and B fields are therefore usual real scalar fields. The expansions of the E and B fields in the harmonics domain is related to the Q and U fields as follows:
\begin{eqnarray}
a^E_{\ell m} = -\frac{{}_2 a_{\ell m} + {}_{-2} a_{\ell m}}{2}\\
a^B_{\ell m} = i \frac{{}_2 a_{\ell m} - {}_{-2} a_{\ell m}}{2}.
\end{eqnarray}
The E and B fields are therefore derived from the Q and U fields by simply applying twice the spin lowering operator to $Q+iU$ and the spin raising operator to $Q-iU$.  Following \citep{starck:pola09}, extending the isotropic and undecimated wavelet transform for E/B polarization fields by constructing formal E and B maps as described in Equation~\ref{eq:EBfield} from the Q and U maps and then applying the isotropic undecimated wavelet transform for each E and B scalar-valued maps independently.

\paragraph{Standard approaches for polarized CMB map estimation:}

\cjb{Performing component separation on a polarization vector field can be performed either in the Q/U coordinate system or the E/B field. So far, most standard methods use a description of the components based on second order statistics. From that respect, the CMB is more naturally described by its E/B power spectra and cross-spectra in the E and B fields. Therefore, the following state-of-the-art methods perform in the E/B fields:
\begin{itemize}
\item{Internal Linear Combination - ILC:} this component separation technique is well-known in the astrophysics community \citep{ILC_eriksen,WMAP9_1,WMAP_Naselsky_Pola}. Essentially, it estimates a CMB map, whether in temperature or polarization, with minimum variance. The Needlet-ILC \citep{2012arXiv1204.0292B} performs in the wavelet domain; it has been applied to the WMAP data and more recently to the Planck PR2 data \citep{PR2_compsep} to provide estimates of the CMB polarized maps. In statistics, this estimation procedure is well known as the BLUE or Best Linear Unbiased Estimator.
\item{Spectral matching ICA - SMICA:} this component separation method enforces the statistical independence of the sought after components - for more details, see \citep{SMICA02,ica:Del2003}. SMICA assumes that each component, whether the CMB or foreground components, can be modeled as a random Gaussian random field with unknown power spectrum. In this setting, the components are estimated by enforcing the contrast between the spectrum of the components in the harmonic domain. An extension of SMICA to polarized data has been proposed in \citep{PolEMICA}. 
\end{itemize}
The main limitations of these two component separation techniques are twofold: i) galactic and extragalactic foregrounds can hardly be modeled as stationary and homogeneous signals and ii) they both rely on second-order statistics to disentangle the components, which is not well-suited to model galactic foregrounds : they are in nature non-stationary and non-Gaussian signals.\\
More recently, grounded on the concept of sparsity, we introduced a new component separation technique coined GMCA \citep{2012arXiv1206.1773B}. This algorithm is based on the sparse modeling of the foregrounds in the wavelet domain. In contrast to the CMB field, the galactic foregrounds are better described in the data domain, and therefore in the Q and U maps. We will therefore rather perform the GMCA algorithm on the observed Q and U maps.\\ \\
In early 2015, the Planck consortium released the first estimates of the polarized CMB maps that have been computed from the Planck PR2 observations \citep{PR2_compsep}. The proposed $4$ maps have been calculated using extensions to polarization of the component separation methods used to process CMB temperature maps \cite{PR1_compsep}. These methods include:
\begin{itemize}
\item {\it Commander: } \citep{Commander} is a Bayesian parametric estimation method that relies on an explicit parameterized modelling of the sky, which includes CMB, synchrotron, and thermal dust emissions.
\item {\it SEVEM: } \citep{WSEVEM} is a template fitting methods that performs in the wavelet domain. For that purpose, templates are derived from differences of input Planck observations and used as templates in a foreground cleaning procedure.
\item{\it NILC: } A version of the Needlet ILC algorithm has been used to derived polarized CMB maps. More precisely, the NILC estimates ILC weights in $13$ wavelet bands (see \citep{PR2_compsep}).
\item{\it SMICA: } A special version of the SMICA algorithm \citep{SMICA02,ica:Del2003} has been used to process the Planck PR2 data. In contrast to the original SMICA algorithm, spectral parameters are fitted for $\ell \leq 50$. For larger multipoles, it is an harmonic ILC.  
\end{itemize}
Component separation precisely amounts to estimating a mixing matrix $\bf A$, as described in Equation~\ref{eq:mixtmodel2}, or equivalently unmixing coefficients for ILC-based methods. As in \citep{PolEMICA,2012arXiv1204.0292B}, these parameters can be estimated jointly for both the E and B fields, which makes perfect sense for the CMB but not necessarily for foreground emissions. Indeed, allowing for different mixing matrices for the Q/U or E/B allows for more degrees of freedom to disentangle between complex contaminants, which are not perfectly described by the linear mixture model of \ref{eq:mixtmodel2}. Subsequently, we propose performing the GMCA algorithm independently in the Q and U fields.}\\

\subsection{Sparse component separation with GMCA}
In contrast to standard component separation methods in cosmology, GMCA \citep{2012arXiv1206.1773B} relies on a different separation principle: it makes profit of the naturally sparse distribution of astrophysical components in the wavelet domain. More precisely, the so-called sparsity property means that most of the energy content of the components is concentrated in a few coefficients. This property is also shared by the CMB since its power spectrum decays rapidly ({\it i.e.} $\ell^2$ at large scale and $\ell^3$ at small scales). This entails that the energy of the CMB will be mainly concentrated in a small number of wavelet coefficients. Components of distinct physical origins are very likely to exhibit different sparsity patterns. The GMCA algorithm assumes that the components are unlikely to share similar high-amplitude wavelet coefficients. Enforcing the sparsity of components in the wavelet domain has indeed been shown to be an efficient separation procedure to achieve a clean, low-foreground, estimation the CMB temperature anisotropies \citep{2012arXiv1206.1773B,PR1_LGMCA}.\\
Let $\boldsymbol{\Phi}$ denotes a wavelet transform. We assume that each source $s_{i}$ can be sparsely represented in ${ \bf \Phi}$; $s_{j}=\alpha_{j}\boldsymbol{\Phi}$. The data $\boldsymbol{X}$ can be written as
\begin{equation}
\boldsymbol{X}=\boldsymbol{A}\mathbf{\alpha}\boldsymbol{\Phi} + {\bf N}\;,
\end{equation}
where $\mathbf{\alpha}$ is an $N_{s}\times T$ matrix whose rows are $\alpha_{j}$.
The GMCA algorithm estimates a mixing matrix $\bf A$ that leads to the sparsest sources $\boldsymbol{S}$. In practice, this is achieved by solving the following optimization problem
\begin{equation}
\label{eq:GMCA_main}
\min_{{\bf A},{\bf \mathbf{\alpha}}}\frac{1}{2}\left\Vert \boldsymbol{X}-\boldsymbol{A}\mathbf{\alpha}\boldsymbol{\Phi}\right\Vert _{F}^{2}+\lambda \left \| \mathbf{\alpha} \right \|_{0}\; .
\end{equation}
The expression $\left \| \mathbf{\alpha} \right \|_{0}$ stands for $\ell_0$ pseudo-norm of $\mathbf{\alpha}$, which counts the number of non-zero entries of $\alpha$. The term ${\bf \left\Vert \boldsymbol{X}\right\Vert }_{\mathrm{F}}= \sqrt { \left(\textrm{trace}(\boldsymbol{X}^{T}\boldsymbol{X})\right)}$ stands for the Frobenius norm.\\ 
The problem in Equation~\ref{eq:GMCA_main} is solved by using an iterative algorithm that optimizes sequentially on the sources ${\bf S} = \mathbf{\alpha} {\bf \Phi}$ and the mixing matrix $\bf A$. According to the model in Equation~\ref{eq:mixtmodel2}, the GMCA algorithm needs to be applied on data that share the same resolution. For that purpose, the Planck frequency channels are first downgraded to a common resolution prior to applying GMCA; estimating a high-resolution CMB map estimate is then carried out by performing GMCA on various subsets of the Planck data. We refer the interested reader to \citep{2012arXiv1206.1773B} for more details about the GMCA algorithm. \\


\paragraph*{Accounting for the variability of the mixture model :}
In polarization as well as in temperature, most foreground emissions, such as the thermal dust and synchrotron emissions, have an electromagnetic spectrum that varies across the sky. From a mathematical point of view this makes the mixing matrix $\bf A$ vary across pixels. To account for the heterogeneity of the mixture, the GMCA algorithm is applied on sky patches in each wavelet band. These different steps gave birth to the LGMCA (Local-GMCA) algorithm, which has been described in details in \citep{2012arXiv1206.1773B}. \\
The LGMCA algorithm has been implemented and evaluated on simulated Planck data in \citep{2012arXiv1206.1773B}. It has been applied to the Planck PR1 data leading to the estimation of a low-foreground full-sky map of the CMB anisotropies \cite{PR1_LGMCA}.  

\subsection{Sparse reconstruction of the CMB polarization maps}
\label{ref:PolGMCA}

\paragraph{Separating non Gaussian and partially correlated components\\}
Whether in polarization or in temperature, foreground removal is an extremely challenging task, particularly :
\begin{itemize}
\item{\bf In the galactic center:} in galactic regions, the foregrounds are, by far, the contaminants with the largest emissivity. The rapid variation of the foreground emission laws at the vicinity of the galactic center makes component separation strenuous, in particular for synchrotron and thermal dust emissions. This has been addressed in \citep{2012arXiv1206.1773B} by localizing the mixture model in the framework of GMCA. However, distinct components are also prone to share some similarities in the galactic region: different components have high emissivity in the same regions. Subsequently, distinct components will exhibit partial correlations in these areas. This phenomenon makes their separation much more straining.\\
\item{\bf At large scales:} in this case -- typically for $\ell < 100$ -- only few spherical harmonics are observed. As a consequence, even theoretically decorrelated components exhibit some experimental correlation, customarily dubbed chance-correlations. The underlying components can therefore be regarded as partially correlated signals.
\end{itemize} 
In both cases, the presence of partial correlations between the components will highly hamper the performances of component separation methods and weaken their ability to efficiently remove galactic foregrounds from the cosmological signal. An extension of the GMCA algorithm \citep{AMCA14} has been recently introduced in the signal processing community to specifically tackle the separation of partially correlated sources. Its main assumption is that partial correlation between components will mainly impact a subset of the sparse decomposition coefficients. The rationale of this algorithm relies on a weighting scheme that aims at penalizing correlated entries, which are the most detrimental for separation. This algorithm first boils down to defining some diagonal weight matrix $\bf Q$ \citep{AMCA14}, and secondly to substitute the minimization problem in Equation~\ref{eq:GMCA_main} with the following one:
\begin{equation}
\min_{{\bf A},{\bf \mathbf{\alpha}}}\frac{1}{2} \mbox{ Trace }\left\{ (\boldsymbol{X}-\boldsymbol{A}\mathbf{\alpha}\boldsymbol{\Phi}) {\bf Q}(\boldsymbol{X}-\boldsymbol{A}\mathbf{\alpha}\boldsymbol{\Phi})^T\right\}+\lambda \left \| \mathbf{\alpha} \right \|_{0}\:.
\end{equation}
In \citep{AMCA14}, it has been shown that partially correlated samples are related to non-sparse columns of the source matrix $\bf S$. Based on this relationship, it adaptively updates the weight matrix with respect to the estimated sources during the iterations of the algorithm.\\
This algorithm has been shown to be very effective to disentangle partially correlated sources. For more details, we refer the interested reader to \citep{AMCA14}.\\

\begin{figure*}[htb]
	\centerline{\includegraphics [scale=0.17]{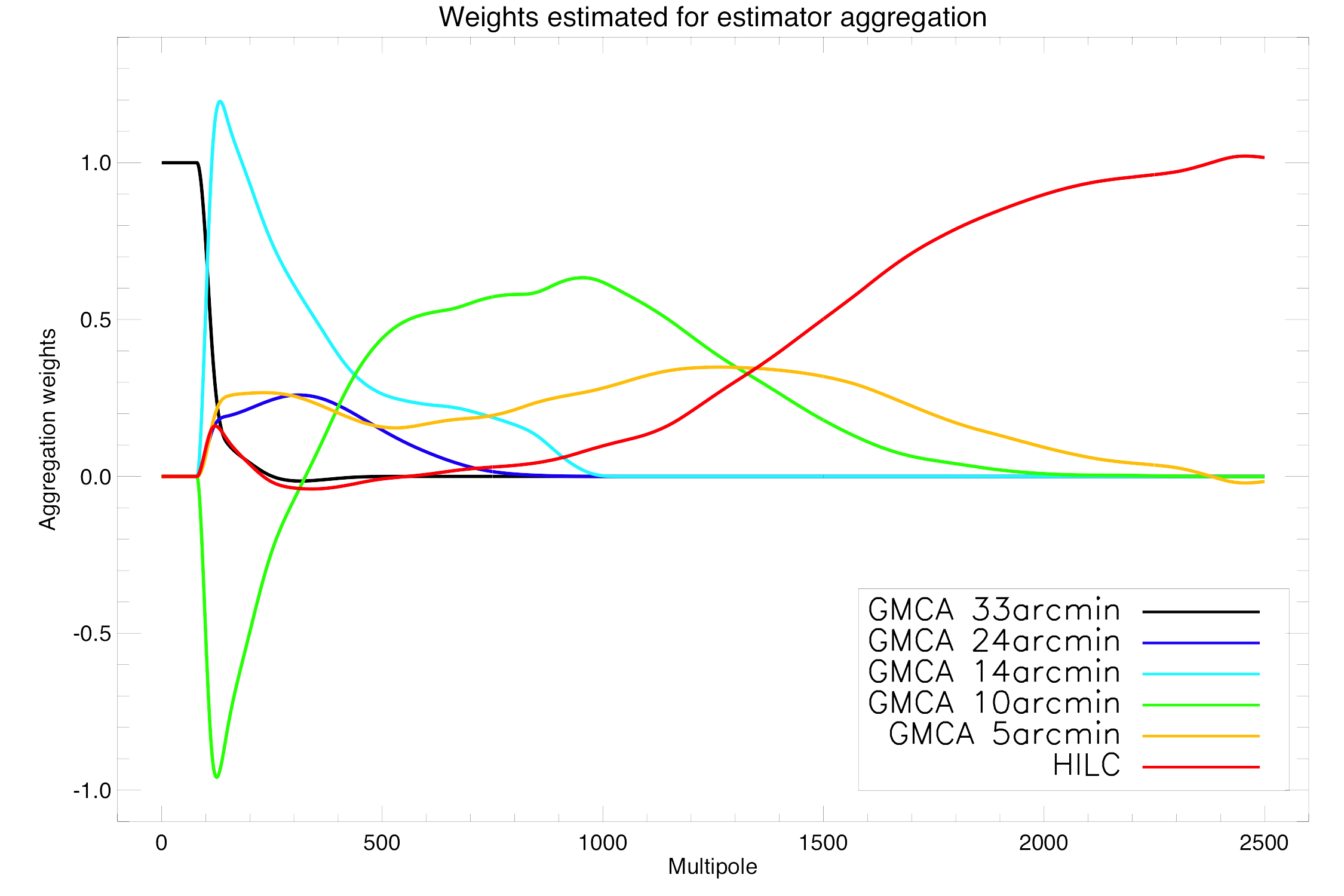}}
	\caption{Weights computed by estimator aggregation.}
	\label{fig_weights_comet}
\end{figure*}

\paragraph{Improving the noise/foreground residual trade-off by combining estimators\\}
Whether in temperature or in polarization, the accurate estimate of CMB anisotropies requires solving somehow antagonistic signal processing issues:
\begin{itemize}
\item{\bf Low noise contamination :} polarized CMB data are highly dominated by instrumental noise, by several orders of magnitude. An effective CMB recovery method should therefore be able to deliver polarized CMB map estimates with low noise contamination. Since instrumental noise in Planck-like data is generally approximated by a stationary Gaussian random field. Hence methods based on second-order statistics (SOS) in the harmonic domain, such Harmonic ILC (HILC - \citep{WMAP_Naselsky_Pola}), are very efficient at small scales where the noise is dominating.\\
\item{\bf Low foreground residuals :} foregrounds, and especially galactic foregrounds, are non-Gaussian and non-stationary signals. As a consequence, component separation methods that rely on SOS, and particularly in the Harmonic domain, are not well suited to accurately remove foreground components. In \citep{2012arXiv1206.1773B}, we emphasized that the use of sparsity in spatially localized signal representations, such as wavelets, provides an effective strategy to estimate a low-foreground CMB map.
\end{itemize}
These two points underlines that component separation for CMB polarization data differs from the case of temperature where the noise is not as large compared to the level of foreground components.
Neither SOS-based nor sparsity-based techniques are perfect candidates to estimate low-foreground and low-noise CMB polarization maps. Subsequently, we rather propose to combine estimates delivered by SOS-based and sparsity-based component separation methods. This aggregation of complementary estimators is inspired by advances in statistics and estimation theory \citep{Yang04}.\\
For the sake of generality, we assume that we have access to $P$ unbiased estimators of the CMB $\{c_i\}_{i=1,\cdots,P}$, which basically differ from each other by their noise contamination and level of foreground residuals. The goal of estimator aggregation is to find some weights $\{\pi_i\}_{i = 1,\cdots,P}$ so that the combined estimator is expressed as a linear combination of these estimators:
\begin{equation}
\hat{c} = \sum_{i=1}^P \pi_i c_i
\end{equation} 
The resulting estimator is guaranteed to be unbiased as long as the weights sum up to unity: $\sum_{i=1}^P \pi_i = 1$. Furthermore, the weights are estimated by minimizing some cost function $\mathcal{R}(\pi_1,\cdots,\pi_P)$ :
\begin{equation}
\min_{\{ \pi_i\}_{i=1,\cdots,P}} \mathcal{R}(\pi_1,\cdots,\pi_P) \quad \mbox{ s.t. } \quad \sum_{i=1}^P \pi_i = 1,
\end{equation}
which enforces some desired properties for the combined estimator.\\
Essentially, the proposed estimator aggregation procedure will amount to combining a SOS-based CMB map estimates, delivered by the HILC, with sparsity-based estimates provided by the LGMCA algorithm. The LGMCA algorithm is well suited for accurately removing foreground residuals but does not provide a low-noise estimate. To that respect, estimator combination is mainly implemented so as to minimize the impact of instrumental noise. Since noise is best modeled in the Harmonic domain with second-order statistics, it is therefore natural to combine LGMCA-based estimates with HILC by seeking the weights $\{ \pi_i \}_{i=1,\cdots,P}$ that minimize the variance of the estimated CMB in the harmonic domain.\\
In practice, deriving a combined estimator in the harmonic domain is done by decomposing the harmonic domain in overlapping bins of multipoles of fixed width $\Delta$, which will be indexed by $l$. A set of weights $\{\pi_{i,l}\}_{i=1,\cdots,P}$ is then estimated in each band of multipoles. Consequently, the aggregation of the $P$ estimators $\{c_i\}_{i=1,\cdots,P}$ is carried out by solving the following minimization problem in each band $l$ of multipoles:
\begin{equation}
\min_{\{\pi_{i,l}\}_{i=1,\cdots,P}} \left \| \sum_i \pi_{i,l} b_{i,l} c_{i,l} \right \|_{\ell_2}^2  \quad \mbox{ s.t. } \quad \sum_{i=1}^P \pi_{i,l} = 1,
\end{equation}
where $\| \, . \, \|_{\ell_2}$ denotes the Euclidean norm, the term $c_{i,l}$  stands for the $l$-th harmonic multipole of the $i$-th CMB map, and the values $b_{i,l}$ are the the beam of different maps at the $l$-th harmonic multipole. This estimator is better known as the Best Linear Unbiased Estimator (BLUE) in the statistics community or the ILC in astrophysics. The different weights are then re-combined to provide values for each multipole $\ell$. \\
Interestingly, the LGMCA algorithm amounts to applying the GMCA algorithm on various subsets of observations. This yields various estimates of the CMB maps with various spatial resolutions : estimated maps with the lowest resolution are computed from a larger number of observations and vice-versa. We have pointed out in \citep{PR1_LGMCA} that applying component separation to a large number of observations allows for more degrees of freedom to better provide a clean estimate of the CMB map. The current version of the LGMCA algorithm performs by using a single CMB map estimate in each wavelet band. From the perspective of estimator aggregation, the LGMCA algorithm already combines complementary CMB maps but in a rather {\it suboptimal} way. Again, the framework of estimator aggregation can help  providing a more efficient combination of GMCA-based estimators with a better balance between noise contamination ({\it i.e.} using maps with different spatial resolutions) and foreground residual ({\it i.e.} using maps obtained from various subsets of data).\\
The PolGMCA algorithm therefore i) derives $5$ estimates of the CMB map from various subsets of observations and with different spatial resolutions (see Section~\ref{comp:protocol}), ii) combines these sparsity-based estimates together with the HILC estimates, and iii) it provides an accurate estimation of large-scale multipoles without being highly sensitive to chance-correlations. Conversely, since the estimator aggregation procedure is based on second-order statistics, it may be prone to chance-correlations at large-scale. The aggregation weights are therefore set to zero for $\ell < 100$ with the exception of the LGMCA low-resolution estimator.\\
To illustrate how estimator aggregation performs, Figure~\ref{fig_weights_comet} displays the values of the aggregation weights for each  of the $5$ Q map estimates computed with LGMCA and the HILC map. These results have been obtained by performing the PolGMCA algorithm to the Planck-like PSM simulations described in Section~\ref{sec:results}.\\
It is very important to notice that, as expected, the contribution of each of the maps delivered by the LGMCA algorithm depend directly on their resolution: low-resolution maps have significant weights at large-scales and vice-versa. Moreover, HILC mainly contributes to the final estimate at small scales ($\ell > 700$) and is predominant at the smallest scales ($\ell > 1500$) where noise is very likely the predominant source of contamination. The global evolution of these weights clearly corroborates the original motivation of estimator aggregation : i) give more weight to sparsity-based estimators when non-Gaussian foregrounds are likely the most prominent, and ii) give more weight to the SOS-based estimator when instrumental noise is the main contamination.

\section{Numerical experiments}
\label{sec:results}

\subsection{Simulations and experimental protocol}
\paragraph{Simulations}
By the end of 2015, the Planck satellite will provide the finest observations of the polarized sky in the frequency range $30$ - $353$GHz. We naturally propose evaluating the proposed component separation method with simulated polarized Planck data. For that purpose, we will make use of simulations generated with the Planck Sky Model (PSM)\footnote{For more details about the PSM, we invite the reader to visit the PSM website: \url{http://www.apc.univ-paris7.fr/~delabrou/PSM/psm.html}.} in (\cite{PSM12}). 

\paragraph{PSM simulations}
\label{sec:appendix} 

The PSM includes the best of the current knowledge of most polarized astrophysical signals and foregrounds, as well as simulated instrumental noise and beams. In detail, the simulations were obtained using the publicly available PSM version 1.7.8 with the following:

\begin{itemize}

\item{\bf Frequency channels:} the simulated data are composed of the $7$ LFI and HFI channels at frequency $33,44,70,100,143,217,353$ GHz. The frequency-dependent beams are perfect isotropic Gaussian PSF with FWHM ranging from $5$ arcmin at $217,353$ GHz to $33$ arcmin at $33$GHz. These simulations do not account for the effect of the scanning strategy adopted by full-sky surveys; indeed, these survey strategies are akin to produce non-isotropic and spatially varying beams, which are more elongated in the scanning directions.\\

\item{\bf Instrumental noise:} the sky coverage is not uniform and homogeneous, since some areas are more often observed than others. The noise statistics ({\it i.e.} noise variance map for each frequency channel) are assumed to be known accurately and consistent with the Planck noise variance maps. The simulations account for the inhomogeneity of the noise statistics but not for its correlation along the scanning direction. Noise will be assumed to be uncorrelated. \\

\item{\bf Cosmic microwave background:} the CMB map is a correlated random Gaussian realization entirely characterized by its T, E, B power spectrum and T-E cross-spectrum. In the simulations, the CMB intensity and polarization power spectra and cross-spectra are generated according to the $\Lambda$-CDM model derived from the Planck 2013 results. The simulated CMB is Gaussian, and no non-Gaussianity ({\it e.g.} lensing, ISW, $\mbox{f}_{NL}$) has been added. The simulated polarized CMB maps are free of B-mode. \\

\item{\bf Point sources:} infrared and radio sources were added based on existing catalogs at that time. In the PSM, polarized point sources are attributed a random polarization degree and uniformly random polarization angle.\\

\end{itemize}

\cjb{The galactic polarization model, which are composed of the dust and synchrotron emissions. These emissions are based on the PSM \texttt{mamd2008} model \cite{MAMD08} that is parameterized by the intrinsic polarisation fraction. By default, it is set to $15\%$ to match the WMAP data.\\
More precisely,} 
\begin{itemize}
\item{\bf Polarized dust emissions:}  \cjb{The polarization of galactic thermal dust emission is due to the partial alignment of elongated dust grains with the Galactic magnetic field. It is however poorly known; the current PSM simulations are extrapolation from the thermal dust intensity. The emission of the thermal dust is modeled by a gray-body emission law. PSM simulations has been shown to be in good agreement with the Archeops observations of polarized dust \citep{ArcheopsDust04} and WMAP $94$GHz measurement.} \\
\item{\bf Polarized synchrotron emission:} \cjb{The polarized synchrotron emission is obtained by extrapolating WMAP $23$GHz Q and U maps using the same power law model as for the temperature map. This approach has shown to be consistent with the WMAP polarized synchrotron maps \citep{Gold_WMAP7Templates}.}\\
\end{itemize}

\paragraph{Comparison protocol}
\label{comp:protocol}

So far, the component separation methods used to estimate CMB polarization maps are limited to ILC-based techniques \citep{2012arXiv1204.0292B} and an extension of SMICA to polarization \citep{PolEMICA}. Contrary to LGMCA, none of these codes is publicly available. The recent analysis of the Planck PR1 temperature data \citep{PR1_compsep} revealed that ILC-based component separation techniques, and especially its harmonic variant, provided some of the best results in the noise-dominant regime: the SMICA map has been obtained by applying Harmonic ILC (HILC) for $\ell > 1500$. Hence, it is very likely that similar results will hold true in the case of polarization. Consequently, this section will focus on comparing the CMB map derived from HILC (described in the appendix \ref{sec:HILC}) and the new PolGMCA method, which combines HILC with the updated version of the LGMCA algorithm.\\

\begin{table*}
\begin{center}
\vspace{0.1in}
\begin{tabular}{|c|c|c|c|}
\hline
Band & \# common resolution & \# sources & Nominal patch size \\
\hline
\hline
1  &  33 arcmin & 7  & N/A \\
\hline
2 &  24 arcmin  &  6   & 64 \\
\hline
3 &  14 arcmin  & 5   & 64 \\
\hline
4  &  10 arcmin & 4  & 32 \\
\hline
5  &  5 arcmin & 3  & 16 \\
\hline
\end{tabular}
\vspace{0.1in}
\label{tab:params}
\caption{Parameters of the local and multiscale mixture model used in LGMCA and PolGMCA.}
\end{center}
\end{table*}

The LGMCA algorithm requires defining four parameters: i) the number of sources is set to be equal to the number of channels, ii) the common resolution, and iii) the nominal patch size. All these parameters are given in Table~\ref{tab:params} and are chosen as in \cite{PR1_LGMCA}.\\

\subsection{CMB map estimates}

\begin{figure*}[htb]
\centerline{
\includegraphics [scale=0.13]{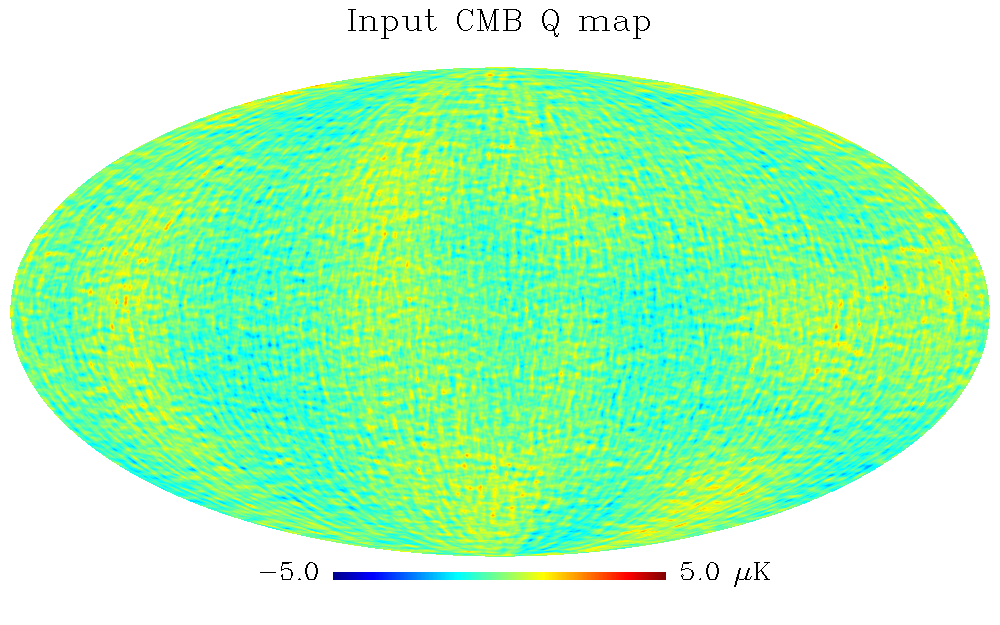}
\includegraphics [scale=0.13]{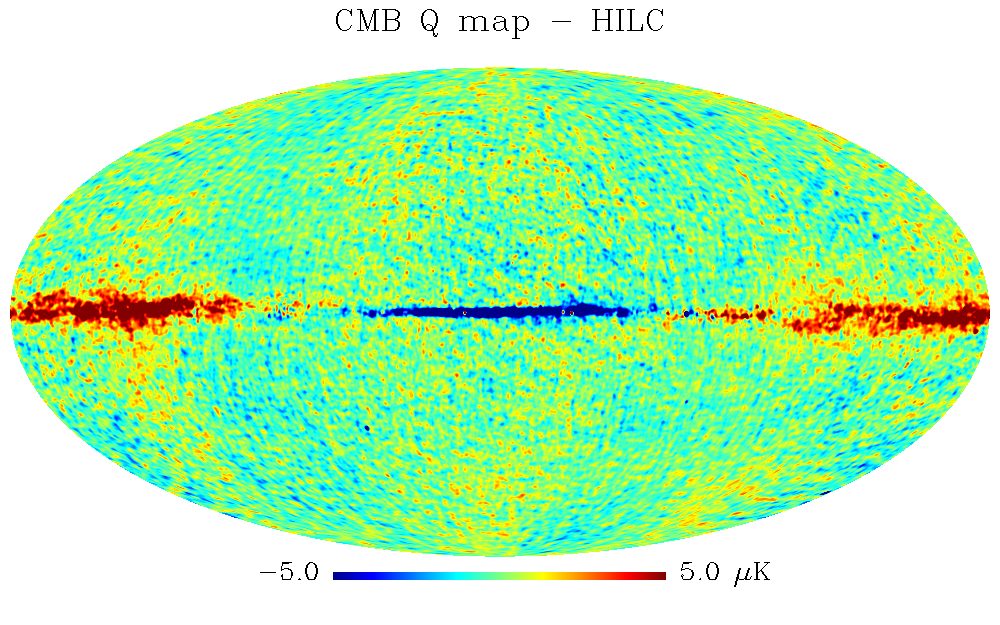}
\includegraphics [scale=0.13]{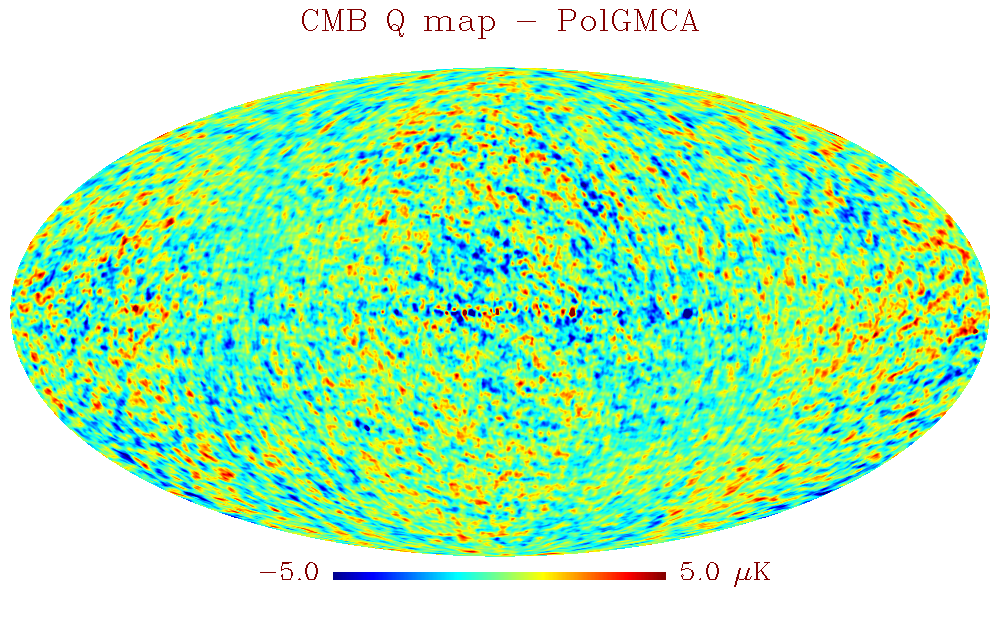}}
\caption{{\bf Estimated CMB Q maps at $1$ degree resolution:} Input CMB (left), HILC (middle), PolGMCA (right).}
\label{fig_map_Qmap}
\end{figure*}

\begin{figure*}[htb]
	\centerline{
		\includegraphics [scale=0.13]{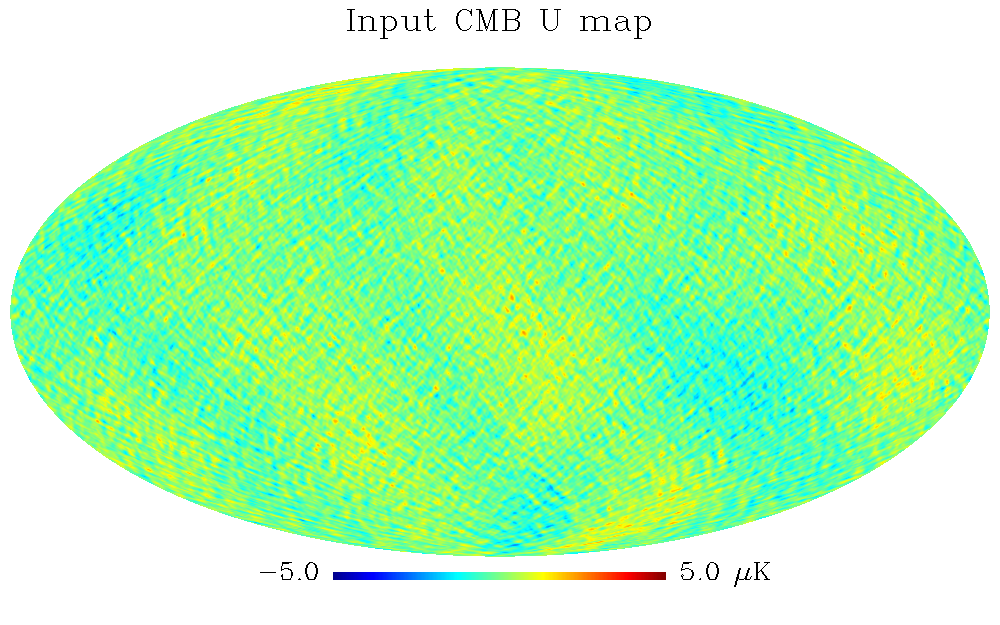}
		\includegraphics [scale=0.13]{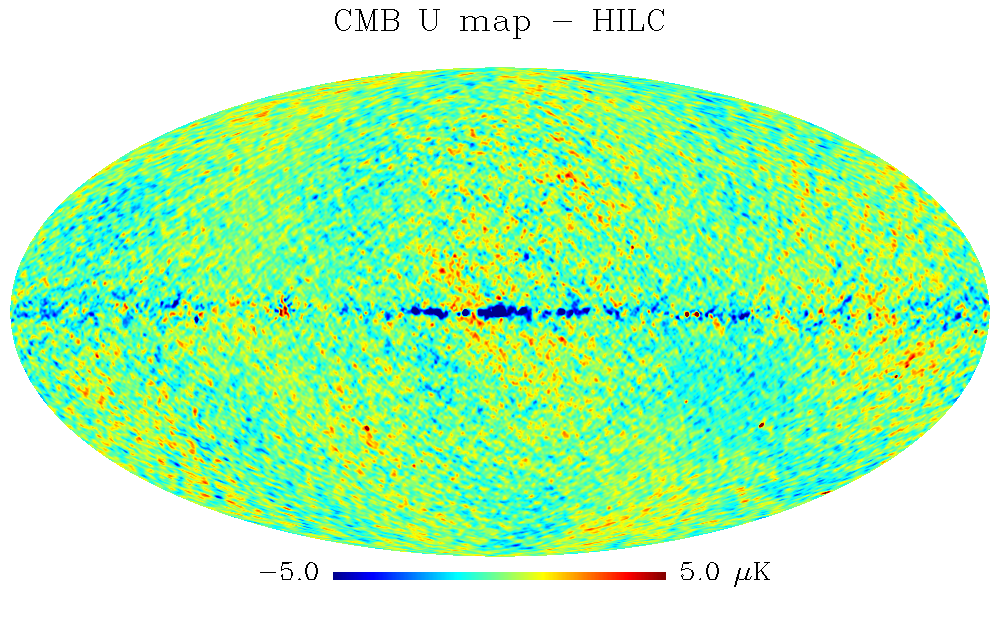}
		\includegraphics [scale=0.13]{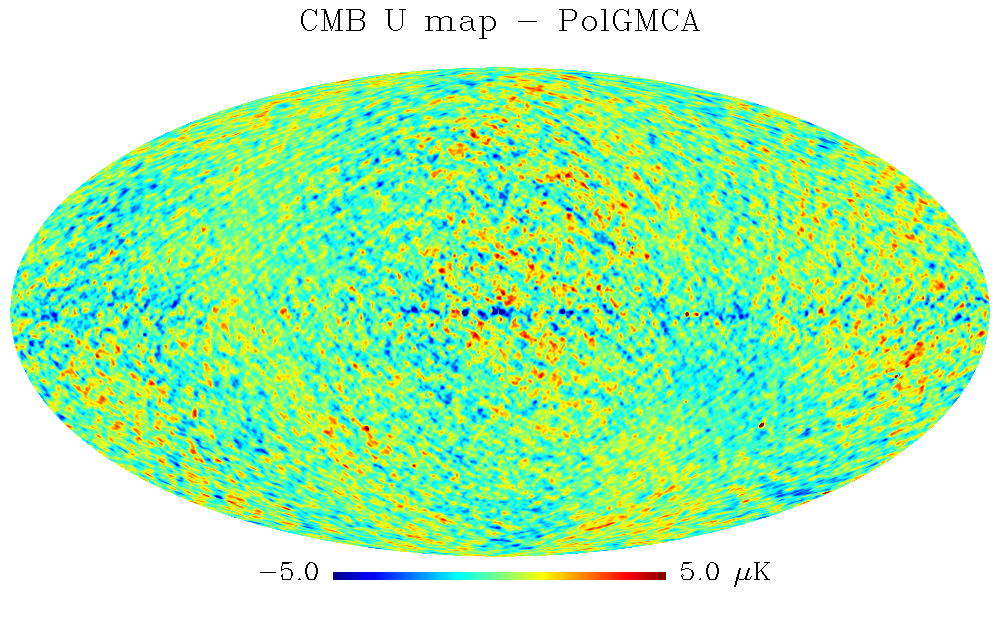}}
	\caption{{\bf Estimated CMB U maps at $1$ degree resolution:} Input CMB (left), HILC (middle), PolGMCA (right).}
	\label{fig_map_Umap}
\end{figure*}

\begin{figure*}[htb]
\centerline{
\includegraphics [scale=0.13]{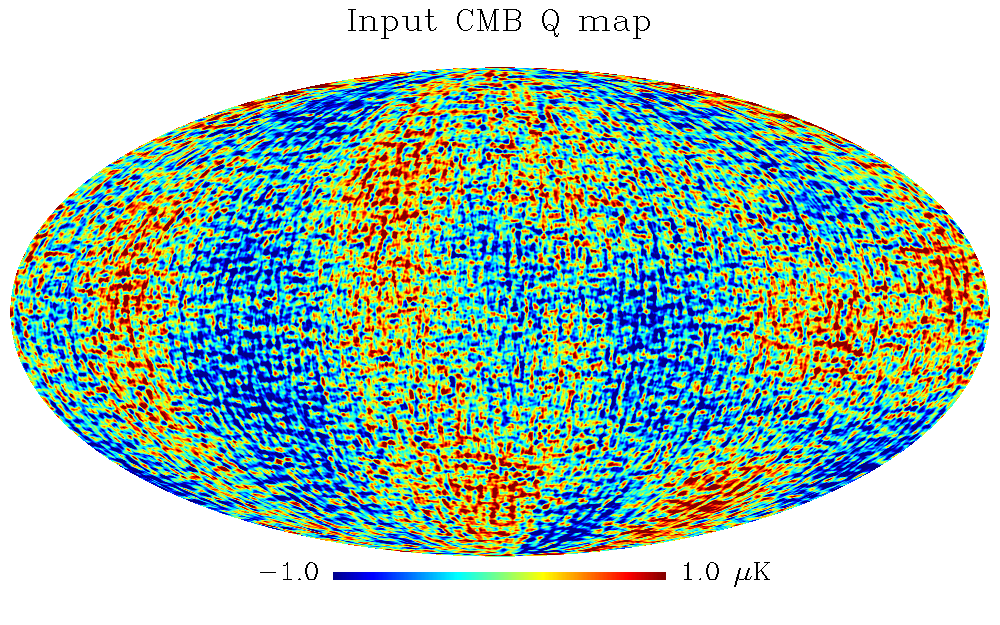}
\includegraphics [scale=0.13]{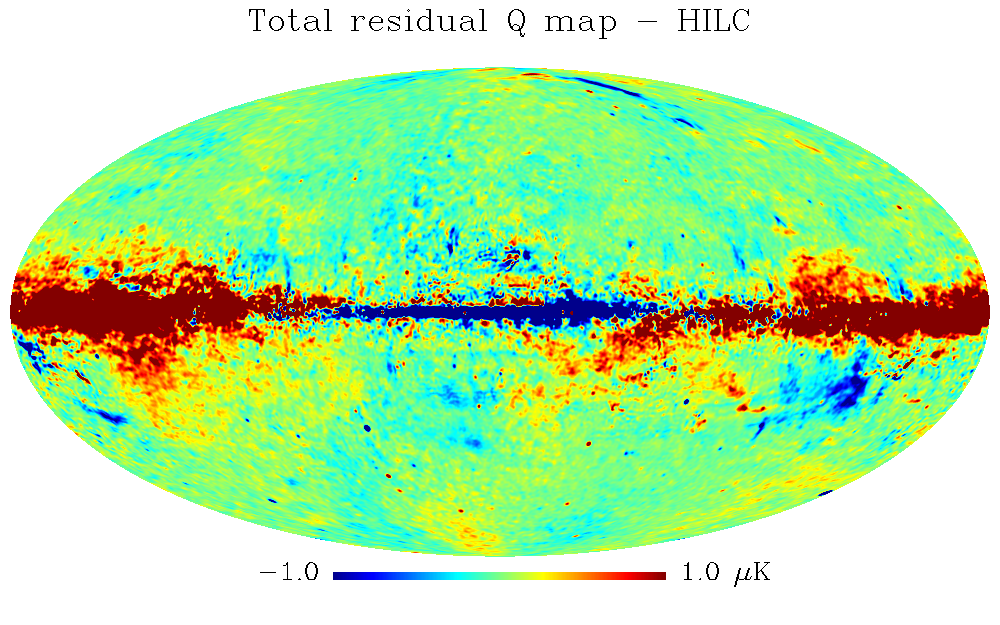}
\includegraphics [scale=0.13]{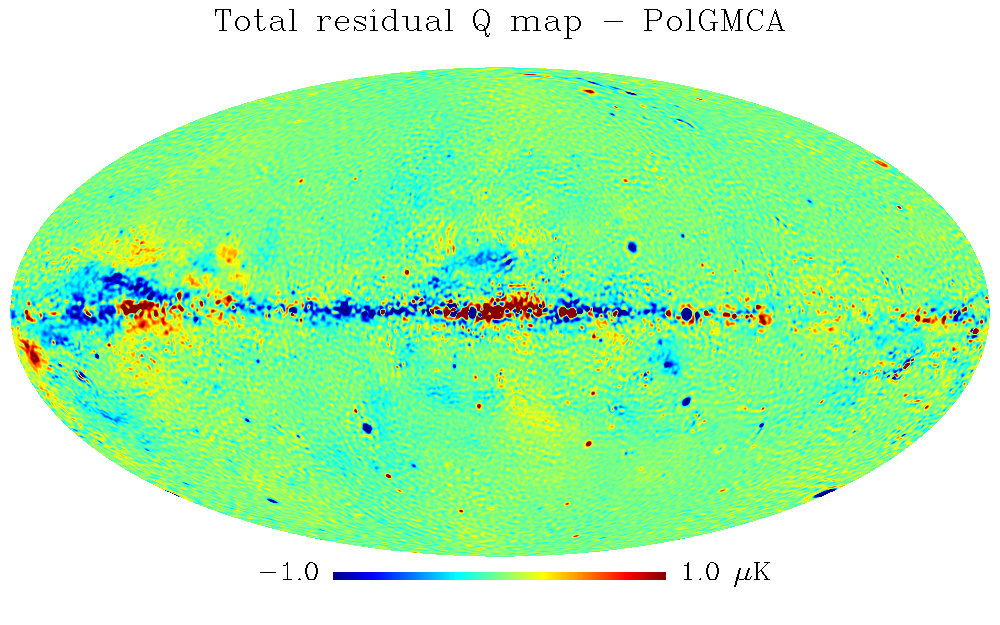}}
\caption{{\bf Foreground residual in the estimated Q maps at $1$ degree resolution:} Input CMB (left), HILC (middle), PolGMCA (right).}
\label{fig_map_Qmap_residual}
\end{figure*}

Figures \ref{fig_map_Qmap} and \ref{fig_map_Umap} display the estimated CMB Q and U maps, which have been computed with HILC and PolGMCA. The maps have been smoothed at the resolution of $1$ degree so as to better highlight the large-scale residuals features and remove high-frequency noise, thus easing visual inspection. These figures first show that the proposed PolGMCA algorithm achieves a much better cleaning of the galactic center; the HILC map clearly contain significant large-scale foreground residual, which are mainly limited to the galactic center. However, while showing less residuals on the galactic center, the PolGMCA map seems to be slightly more contaminated by instrumental noise as seen by seemingly larger background fluctuations. To a lower extent, the same conclusions can be drawn from the estimated U maps features on Figure \ref{fig_map_Umap}.\\
Figure \ref{fig_map_Qmap_residual} shows the total foreground residuals which contaminates the estimated CMB Q maps. These maps are obtained by applying the unmixing parameters to the input foregrounds. These maps have further been smoothed to a common resolution of $1$ degree. Both HILC and PolGMCA visually yield low levels of foreground residuals at large galactic latitudes. As noticed earlier in this paragraph, the PolGMCA algorithm achieves a significantly better foreground cleaning at the vicinity of the galactic center.\\

\begin{figure*}[htb]
	\centerline{
		\includegraphics [scale=0.13]{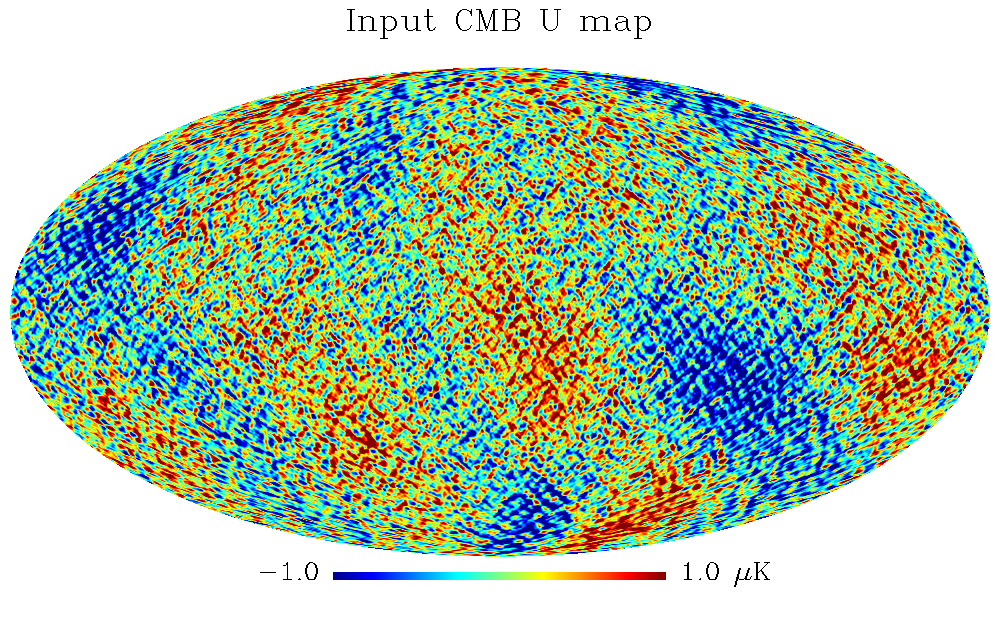}
		\includegraphics [scale=0.13]{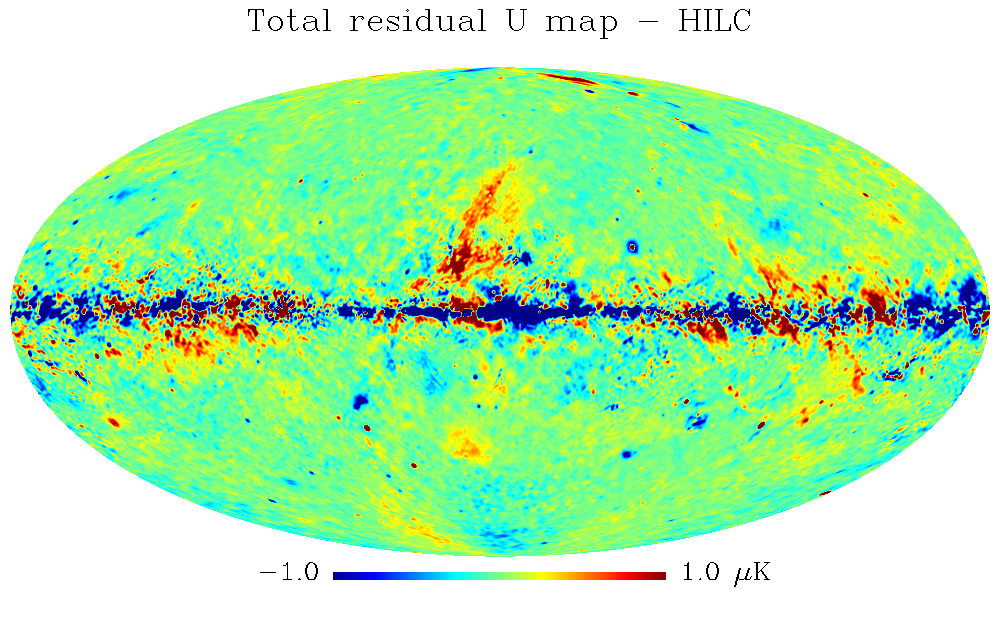}
		\includegraphics [scale=0.13]{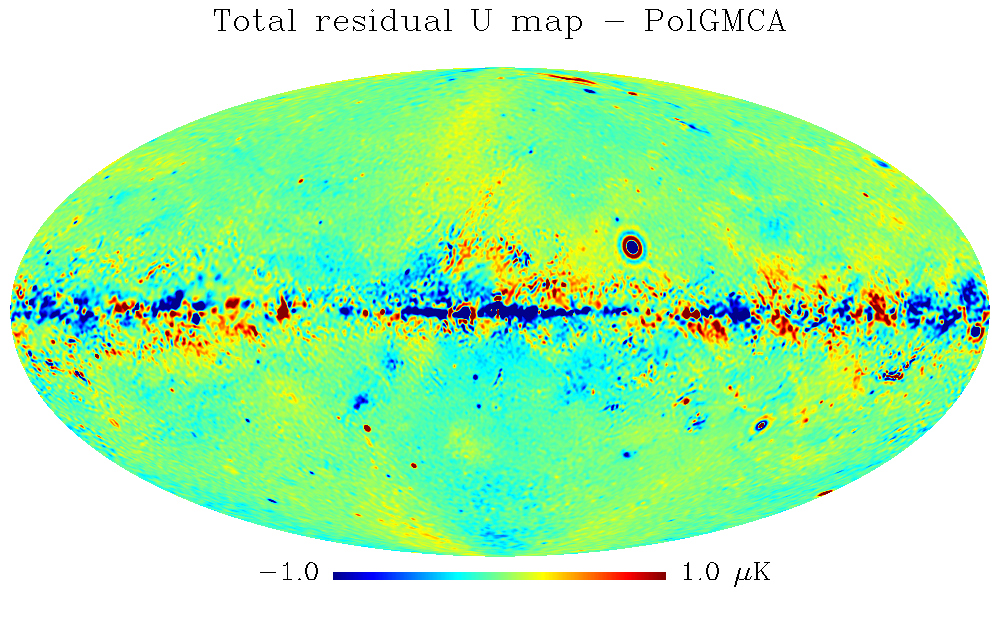}}
	\caption{{\bf Foreground residual in the estimated U maps at $1$ degree resolution:} Input CMB (left), HILC (middle), PolGMCA (right).}
	\label{fig_map_Umap_residual}
\end{figure*}

\paragraph*{Total foreground residual}

\begin{figure*}[htb]
\centerline{
\hbox{
\includegraphics [scale=0.17]{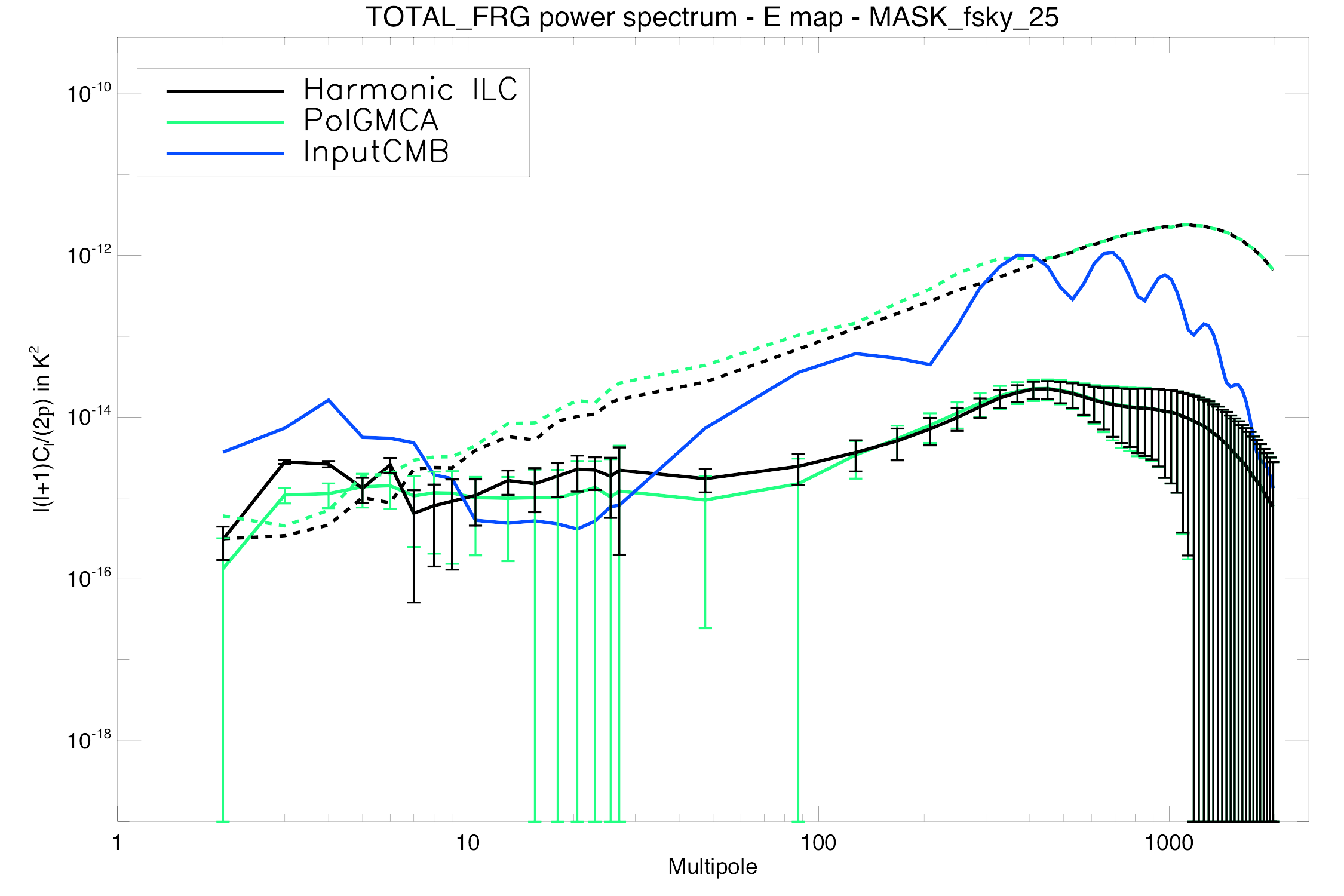}
\includegraphics [scale=0.17]{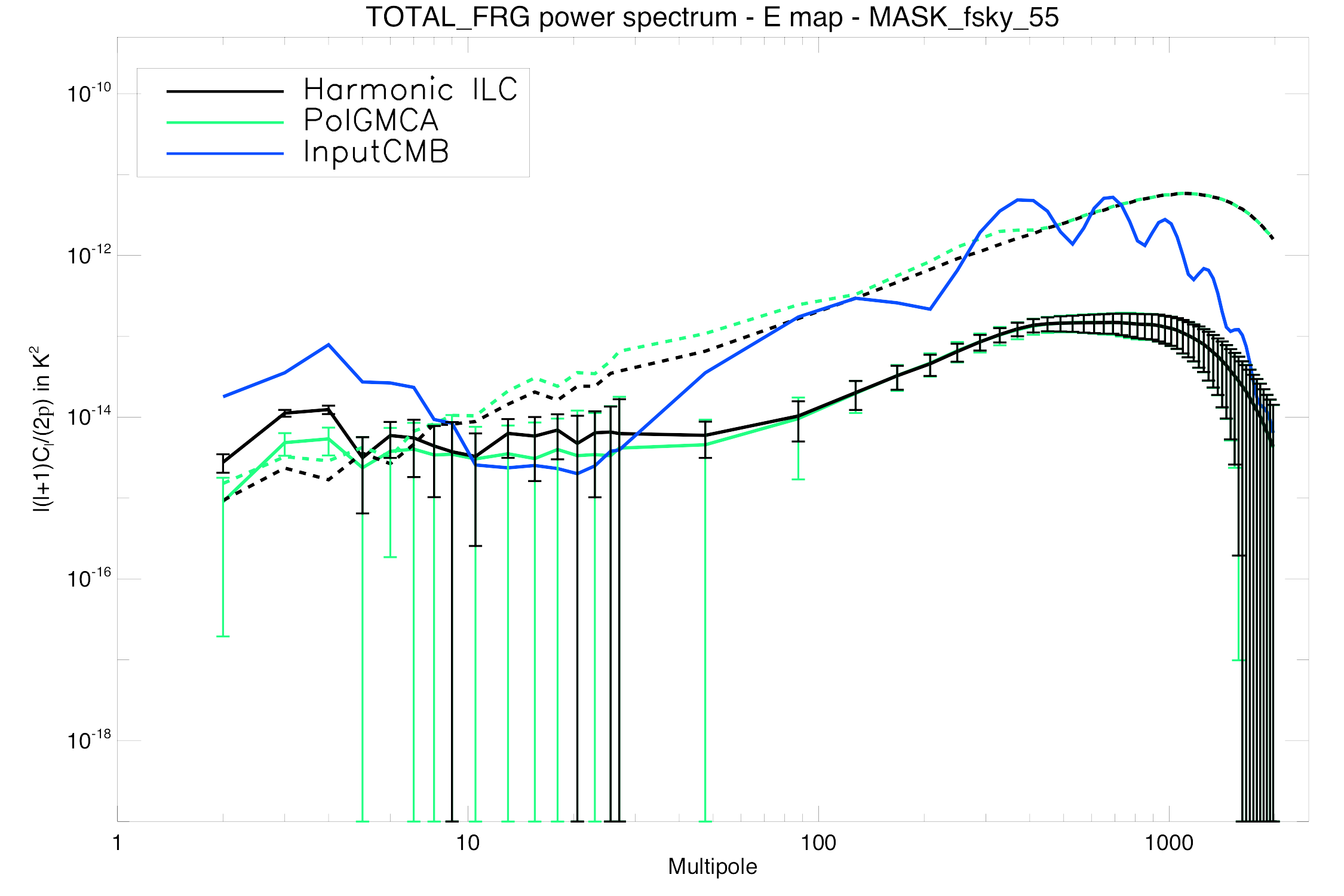}
}}
\vfill
\centerline{
\hbox{
\includegraphics [scale=0.17]{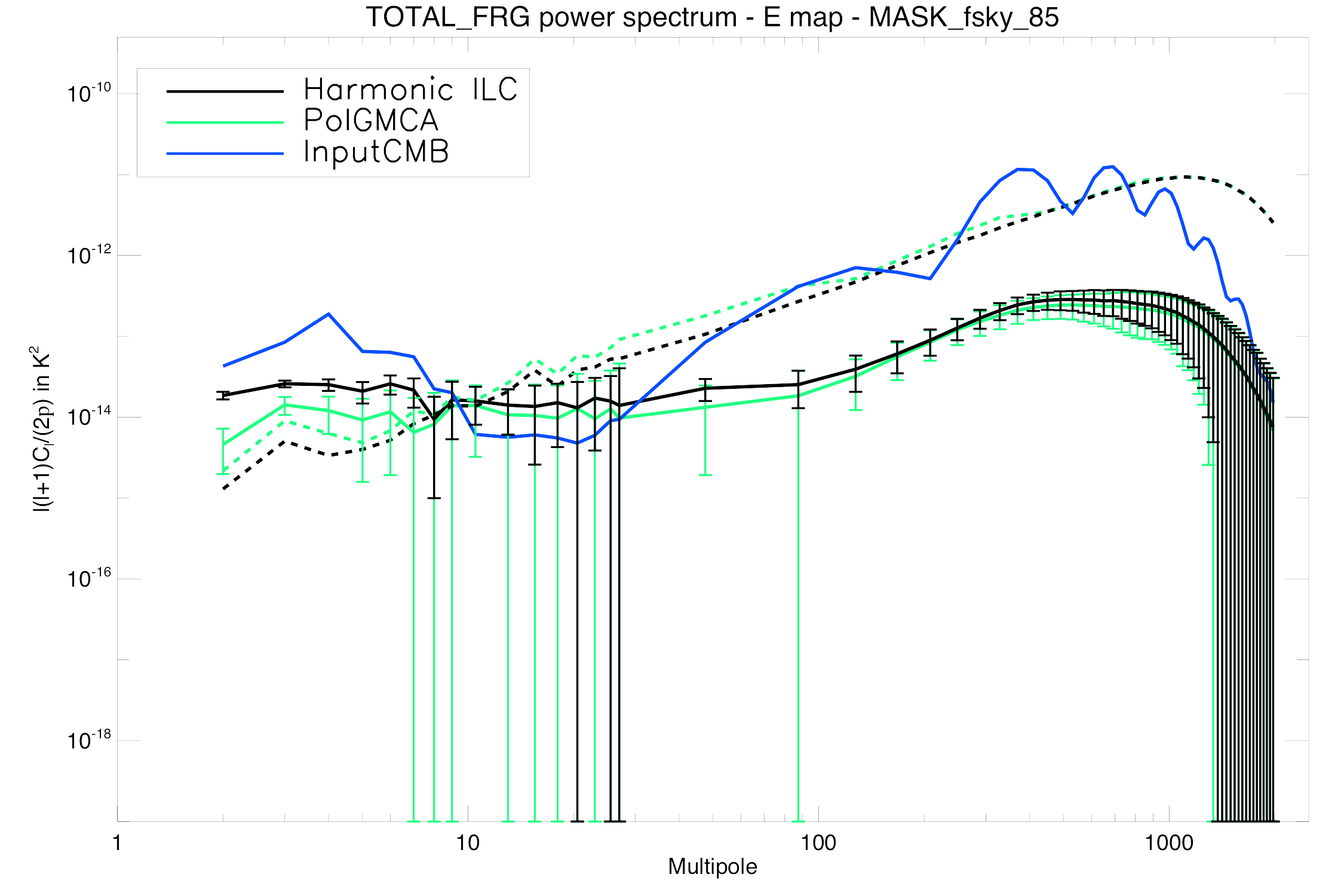}
\includegraphics [scale=0.17]{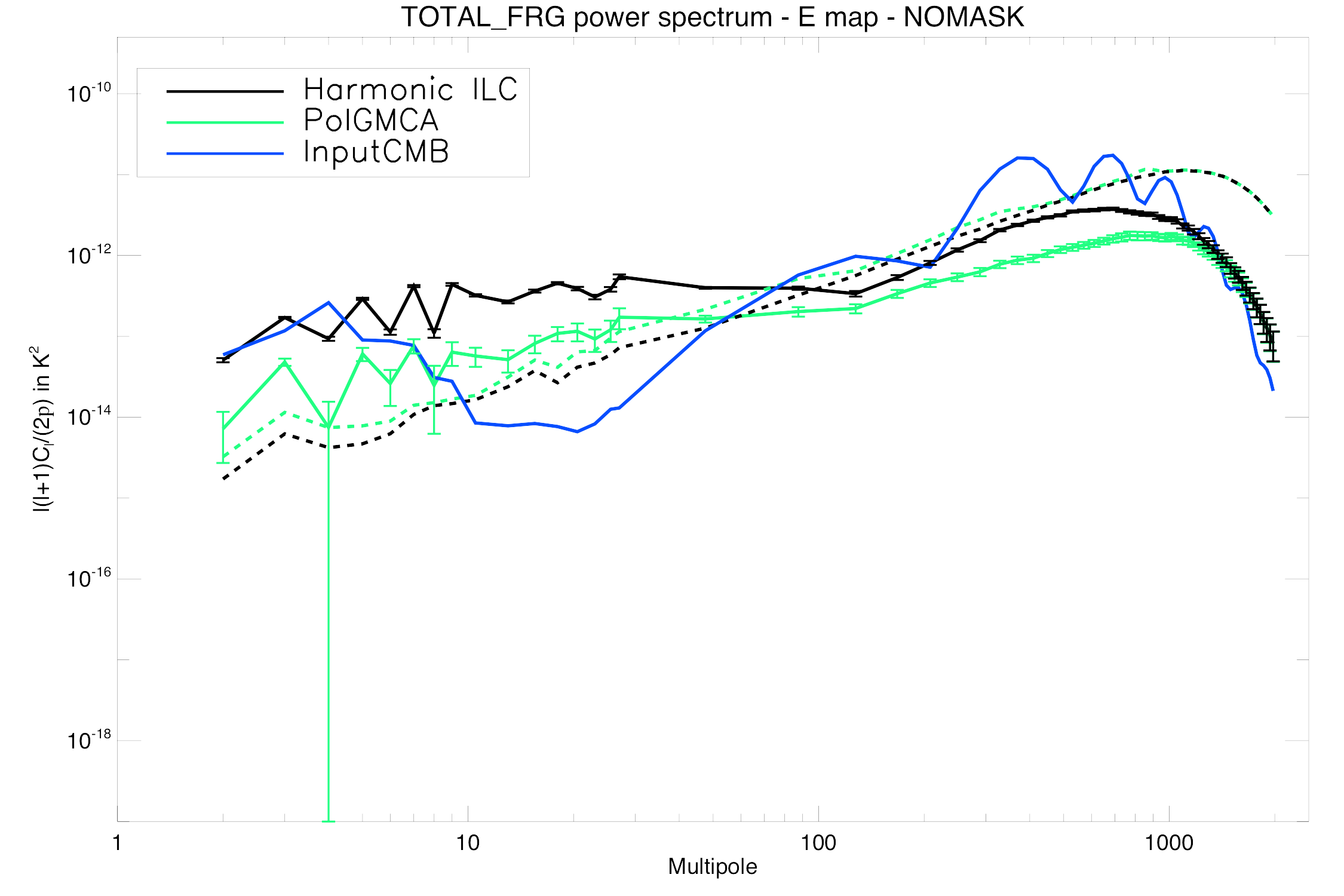}
}}
\caption{{\bf Total foregrounds - top-left :} power spectrum of total foregrounds residual for the E map from a sky coverage of $25\%$. {\bf top-right :} from a sky coverage of $55\%$. {\bf bottom-left :} from a sky coverage of $85\%$. {\bf bottom-right :} full-sky.}
\label{fig_total_frg_Q}
\end{figure*}

\cjb{In this section, we evaluate the quality of estimation of the polarized CMB E maps, which we computed with HILC and PolGMCA. It goes without saying that the purpose of the article is to provide a component separation methods that provides clean, low-foreground, polarized CMB maps at high galactic latitudes as well as on the galactic center. For that purpose, the power spectra of the individual and total foreground residuals have been computed from various values of the sky coverage --$25\%$, $55\%$ and $85\%$  -- and without any masking (full-sky). These masks are common to all the methods; they have been computed based on the foreground residuals of both the two maps, so as not to favor/penalize one of these maps. Since these comparisons are based on simulations, the noise and CMB angular power spectra are the actual spectra of the input simulation; they are therefore not based on Monte-Carlo simulations. Furthermore, these angular power spectra are display with $3\sigma$ error bars based on $100$ noise simulations. The instrumental noise being one of the major contaminant of the estimated polarized CMB maps, the uncertainties with respect to noise allow to assess quite accurately the statistical significance of the foregrounds residuals.}\\
Figure~\ref{fig_total_frg_Q} displays the power spectrum of the total foreground residual ({\it i.e.} synchrotron, thermal dust and point sources) for HILC and PolGMCA for various values of the sky coverage. At large-scale, for $\ell < 100$, the PolGMCA algorithm provides a  lower level of foreground residuals whether it is at large galactic latitudes (sky coverage of $25\%$) or on the galactic center (full-sky estimate). This enhancement is very likely the consequence of the new separation mechanism described in Section~\ref{ref:PolGMCA}, which makes the estimation of the E CMB map less sensitive of chance-correlations, but at the cost of an increase of the noise level. \\
At intermediate scales -- for $100 < \ell < 600$ -- PolGMCA and HILC provide very similar results on $55\%$ of the sky. HILC seem to perform slightly better for high galactic latitudes. Both methods yield estimates with very low foreground residuals at high galactic latitudes, which are about one order of magnitude lower than both the noise and the cosmological signal. For larger sky coverage, the level of foreground residuals rapidly increases in the CMB E map delivered by the HILC algorithm while PolGMCA produces a clean full-sky estimate of the E map at intermediate scale. At smaller scales -- $\ell > 600 $ -- PolGMCA performs better than HILC with similar noise level but lower foreground residuals. Interestingly, at very large-scales -- $\ell < 10 $ -- the level of foregrounds is significantly lower than the level of CMB; this suggests that large-scale anomalies in polarization could also be carried out without masking, similarly to what has been done for the temperature large scale anomalies studies \citep{PR1_LGMCA_anomalies}. This result makes PolGMCA a good candidate to analyze the large-scale anomalies of the CMB map.


\cjb{Figure \ref{fig_total_frg_Q} displays $3\sigma$ error bars based on $100$ noise simulations, which first reveal that outside the galactic ({\it i.e.} for masks larger than $85 \%$) the discrepancy between these two methods is not significant with respect to noise uncertainty. In contrast, when no mask is applied, the improvements led by the PolGMCA algorithm are statistically significant. This highlights the very good performances of the PolGMCA algorithm at the vicinity of the galactic center.}


\paragraph*{Individual foreground residuals}

\begin{figure*}[htb]
\centerline{
\hbox{
\includegraphics [scale=0.17]{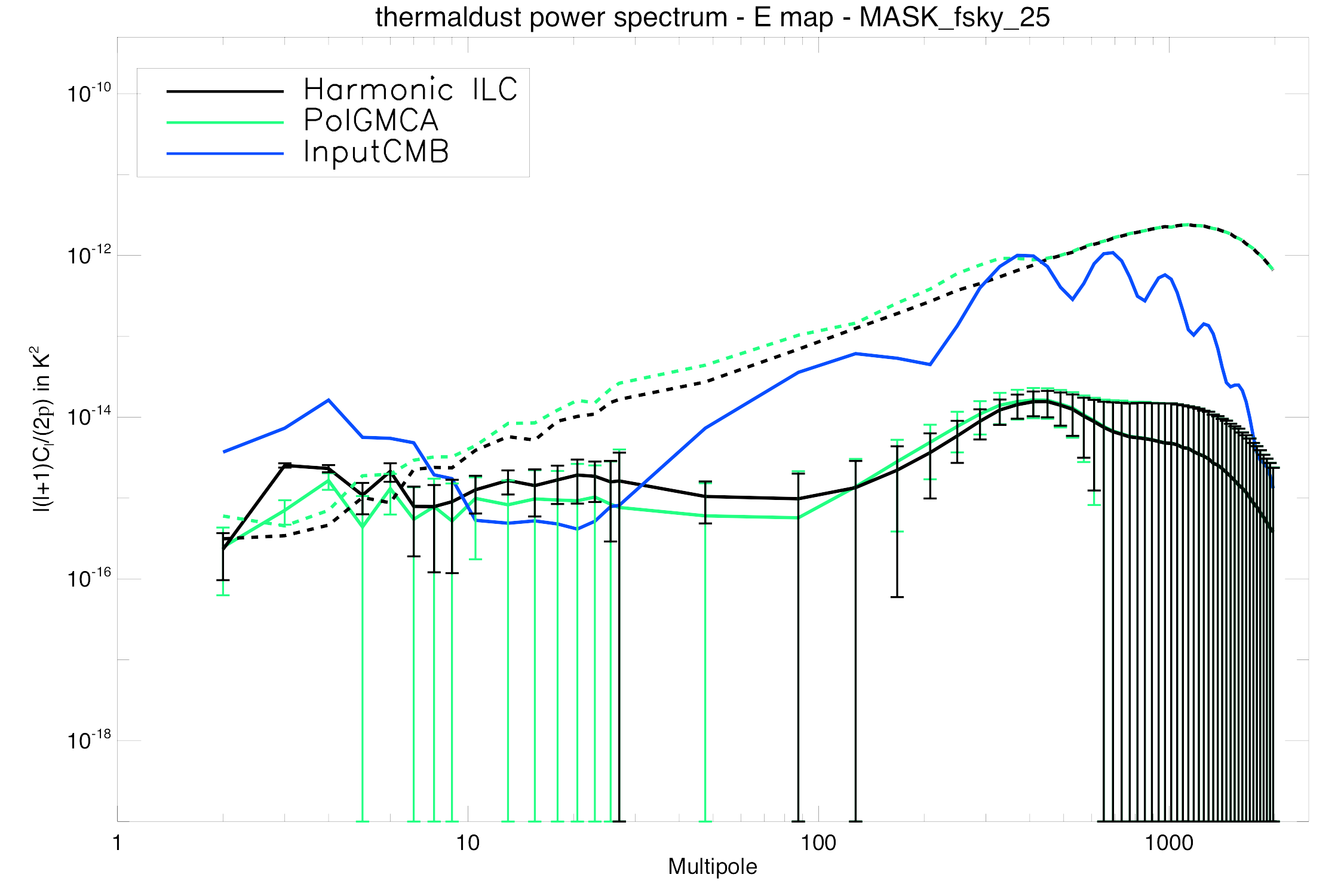}
\includegraphics [scale=0.17]{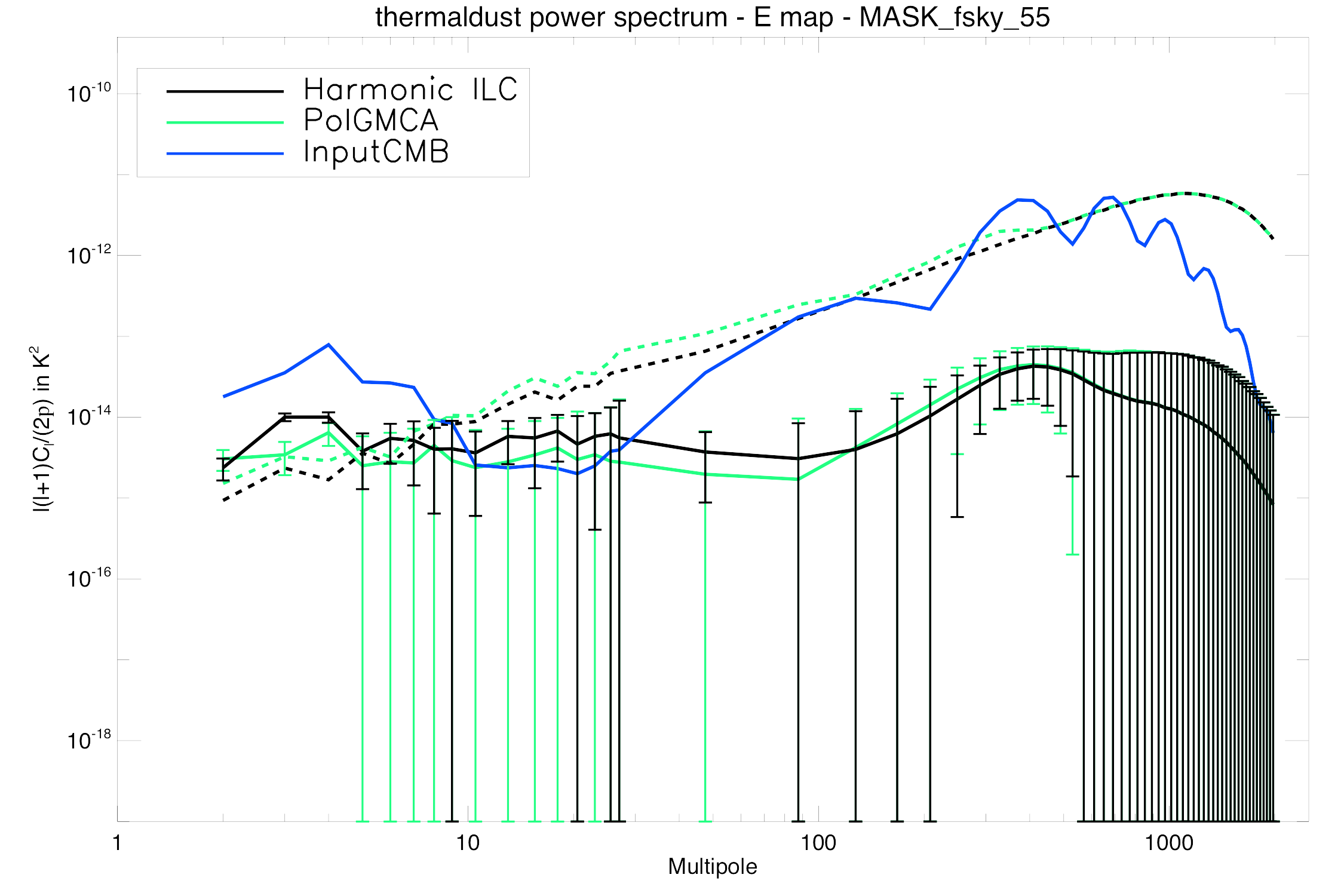}
}}
\vfill
\centerline{
\hbox{
\includegraphics [scale=0.17]{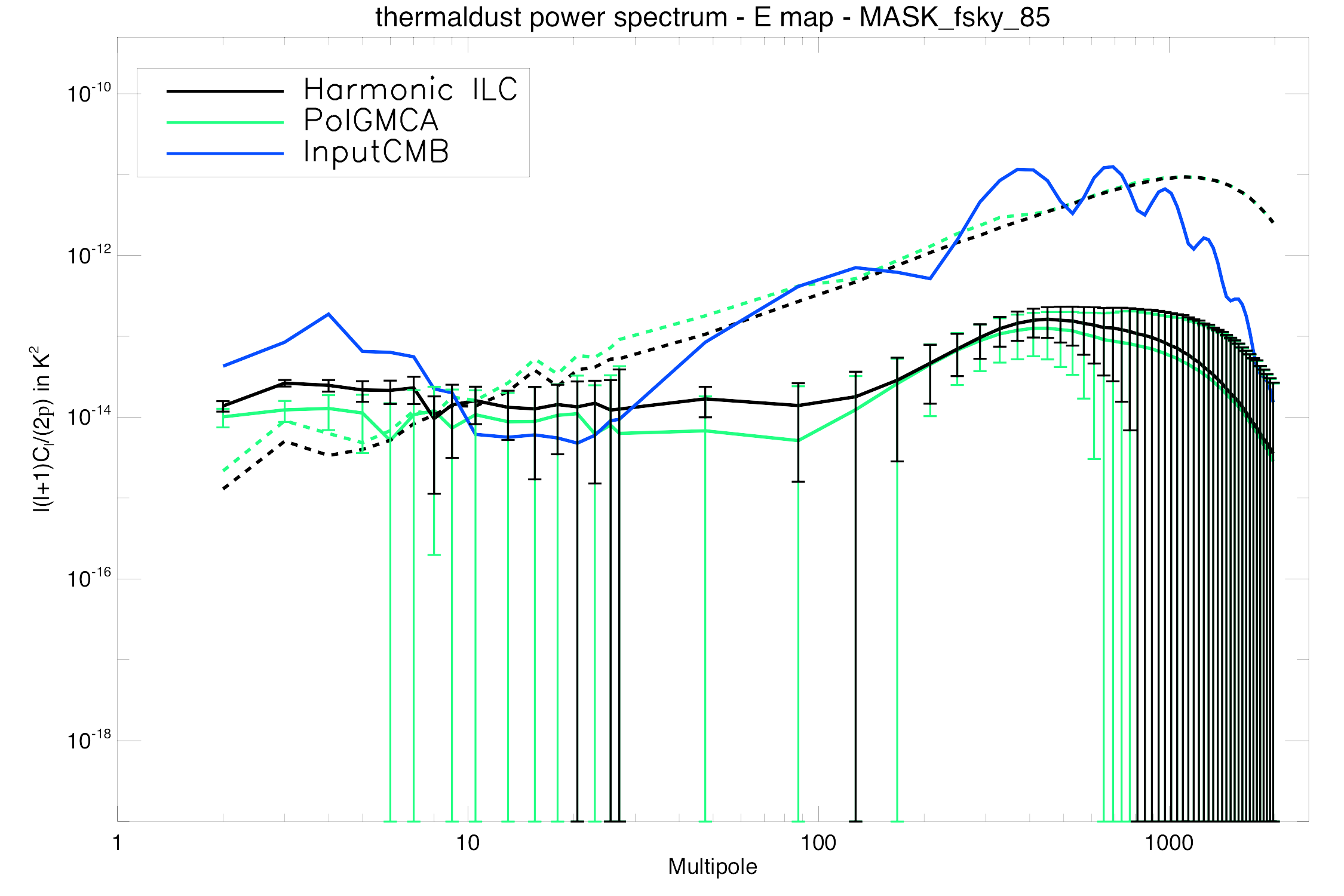}
\includegraphics [scale=0.17]{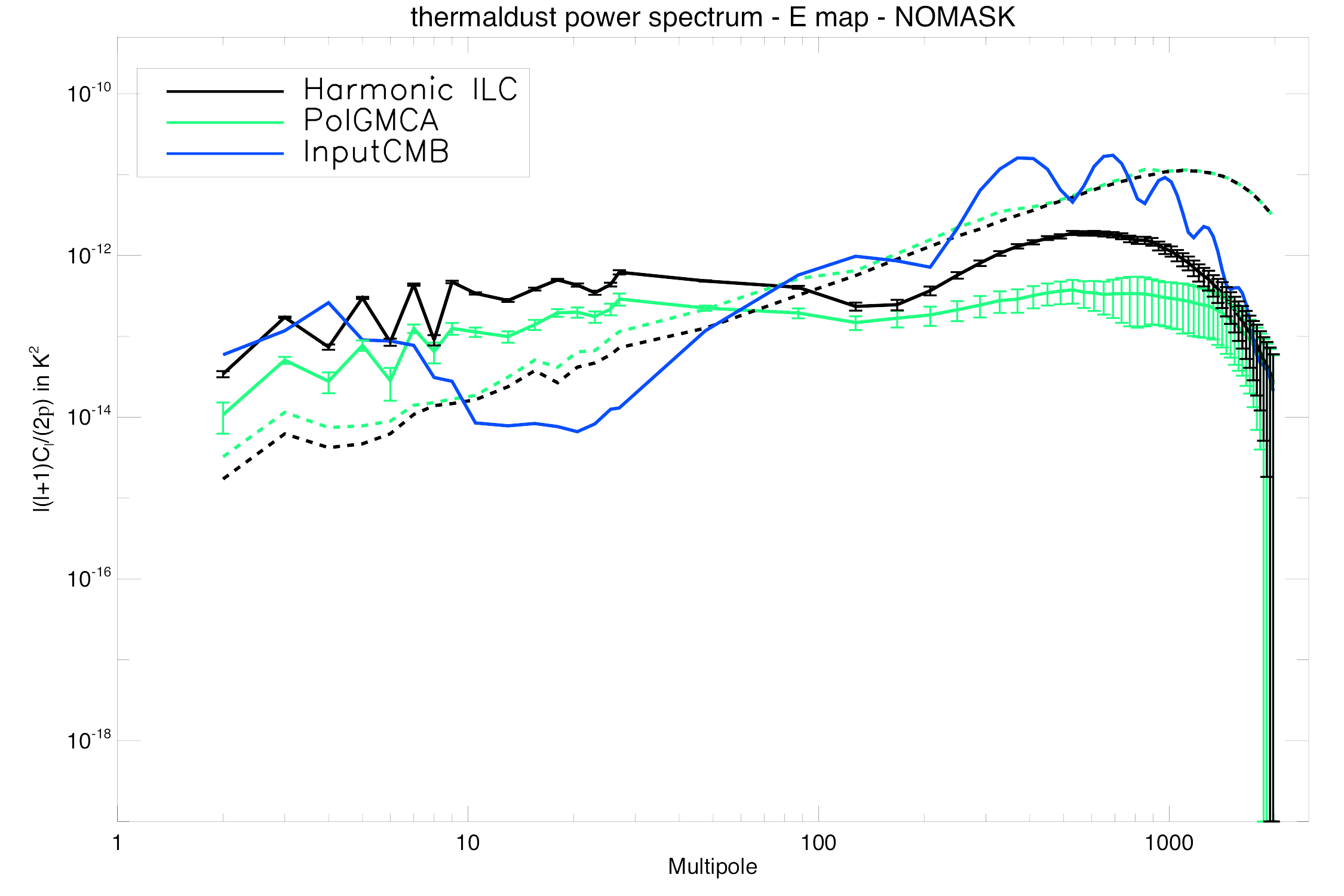}
}}
\caption{{\bf Thermal dust - top-left :} power spectrum of the thermal dust residual in the estimated E map from a sky coverage of $25\%$. {\bf top-right :} from a sky coverage of $55\%$. {\bf bottom-left :} from a sky coverage of $85\%$. {\bf bottom-right :} full-sky.}
\label{fig_tdust_Q}
\end{figure*}

\begin{figure*}[htb]
\centerline{
\hbox{
\includegraphics [scale=0.17]{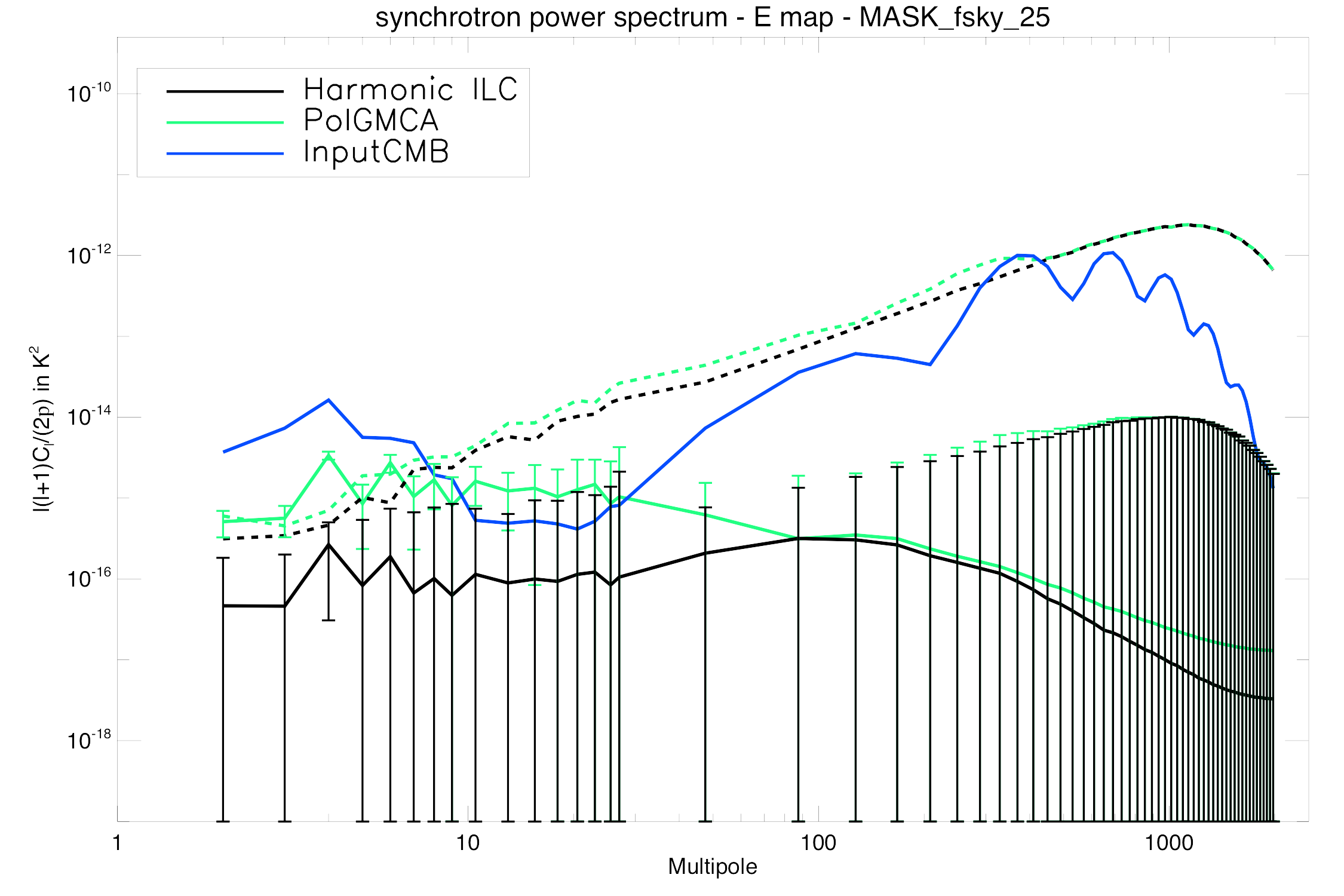}
\includegraphics [scale=0.17]{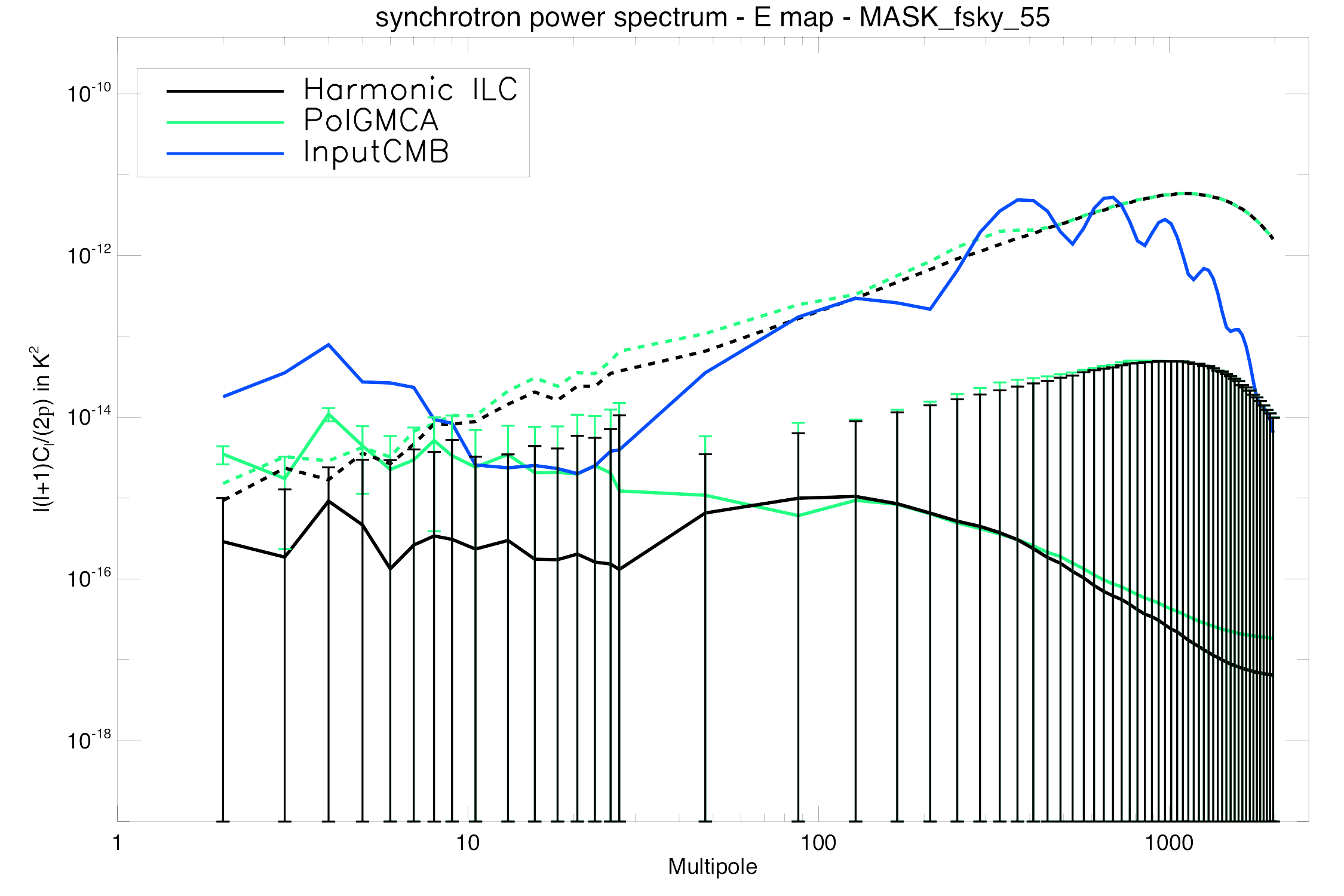}
}}
\vfill
\centerline{
\hbox{
\includegraphics [scale=0.17]{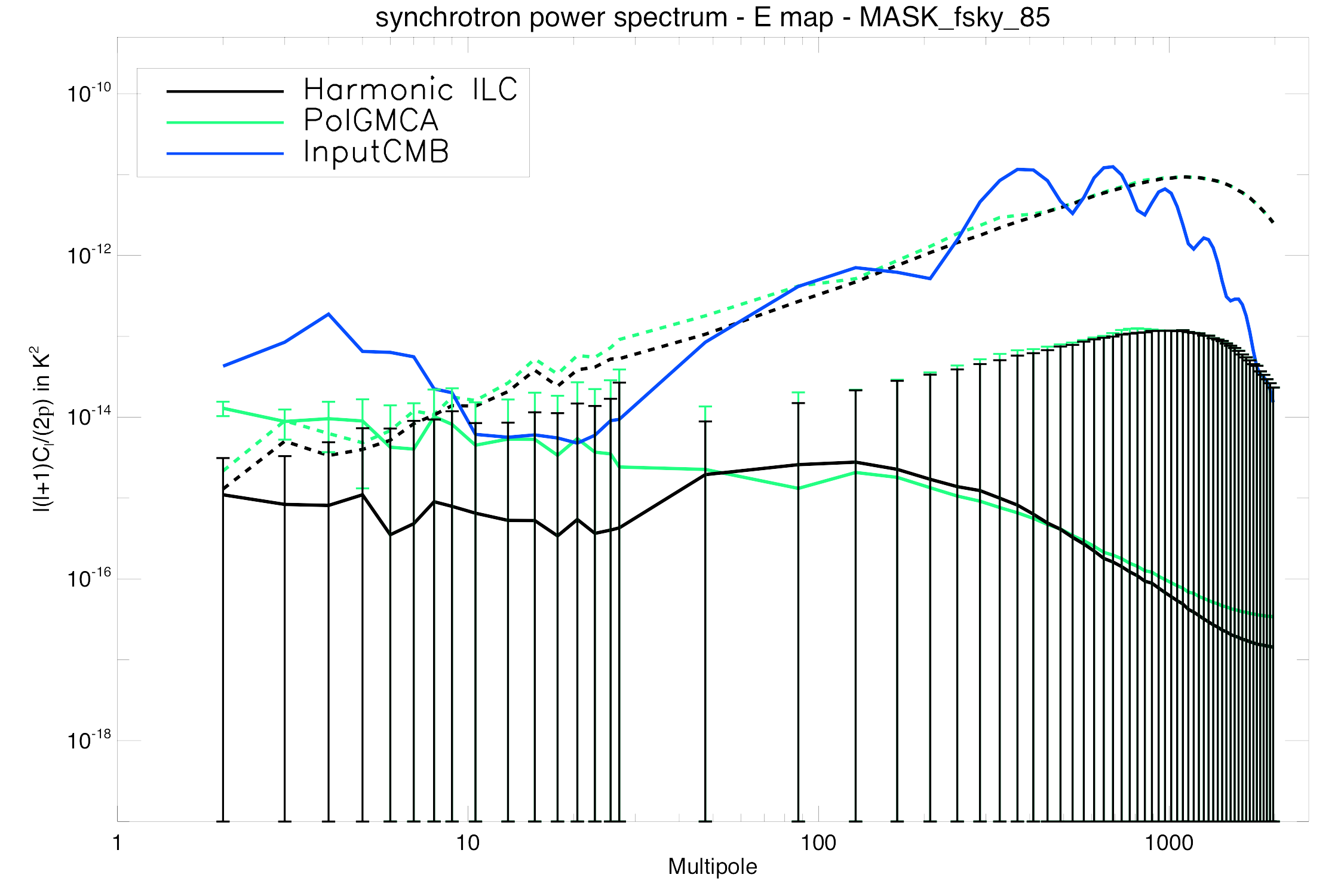}
\includegraphics [scale=0.17]{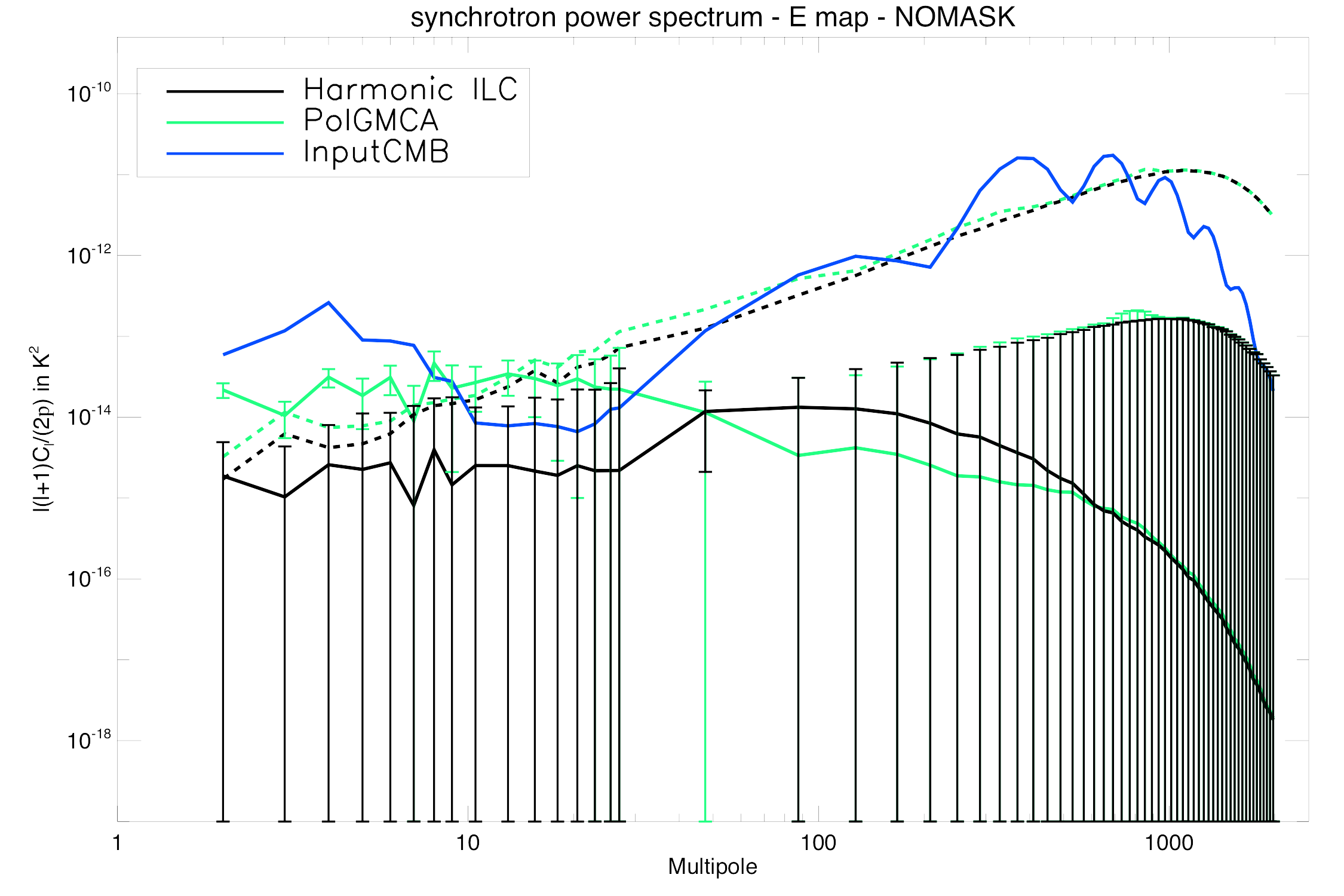}
}}
\caption{{\bf Synchrotron - top-left :} power spectrum of the synchrotron residual in the estimated E map from a sky coverage of $25\%$. {\bf top-right :} from a sky coverage of $55\%$. {\bf bottom-left :} from a sky coverage of $85\%$. {\bf bottom-right :} full-sky.}
\label{fig_sync_Q}
\end{figure*}

\begin{figure*}[htb]
\centerline{
\hbox{
\includegraphics [scale=0.17]{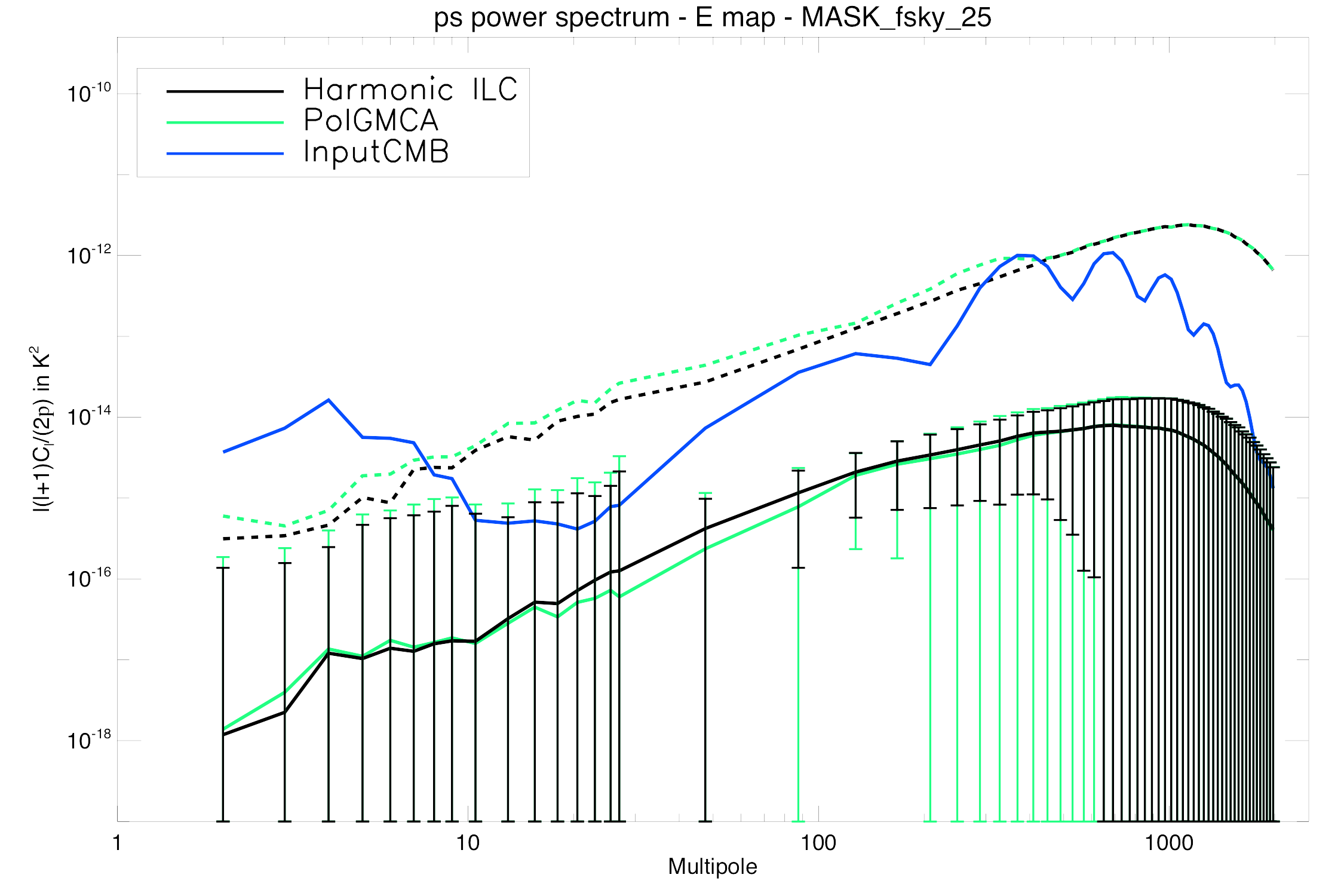}
\includegraphics [scale=0.17]{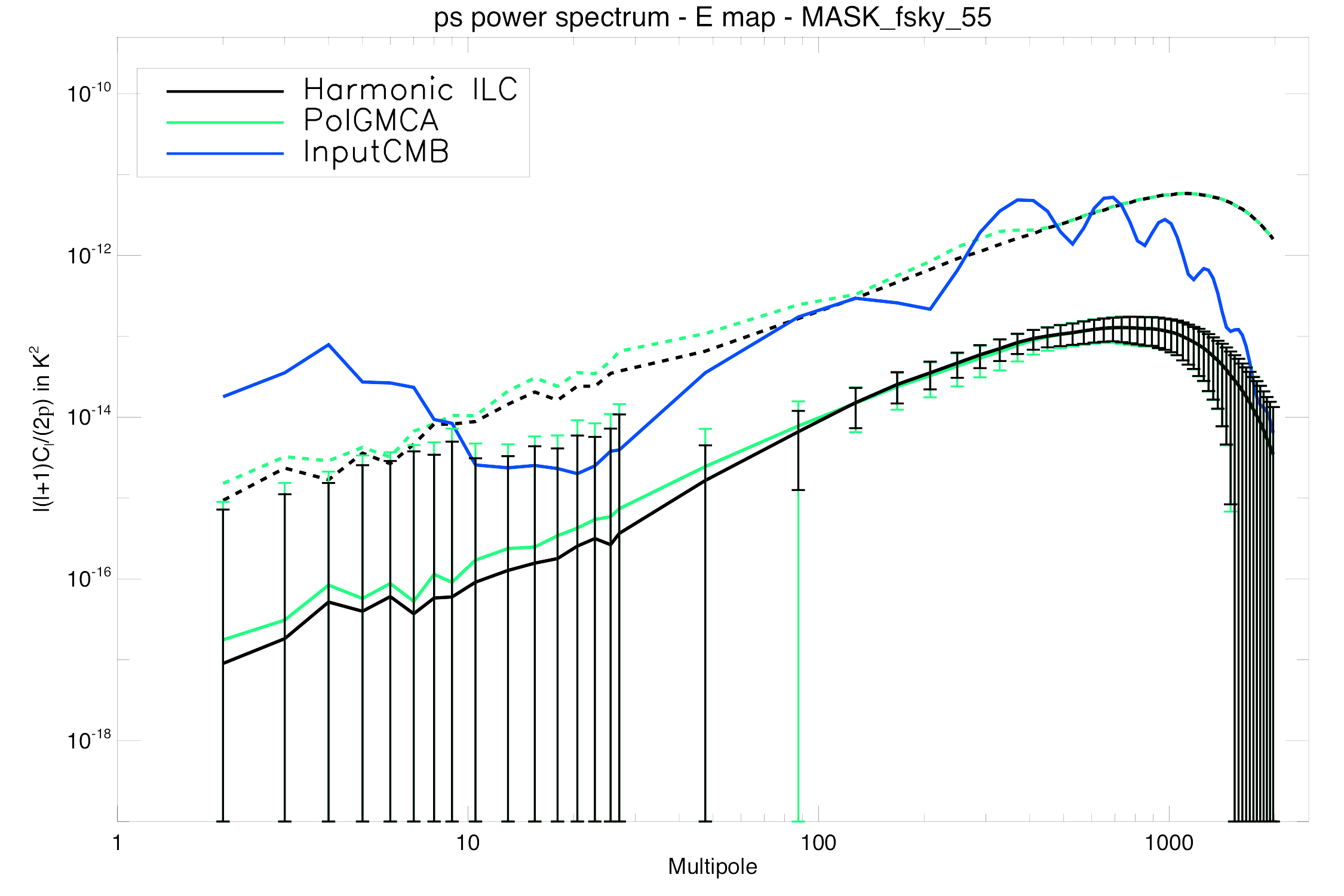}
}}==
\vfill
\centerline{
\hbox{
\includegraphics [scale=0.17]{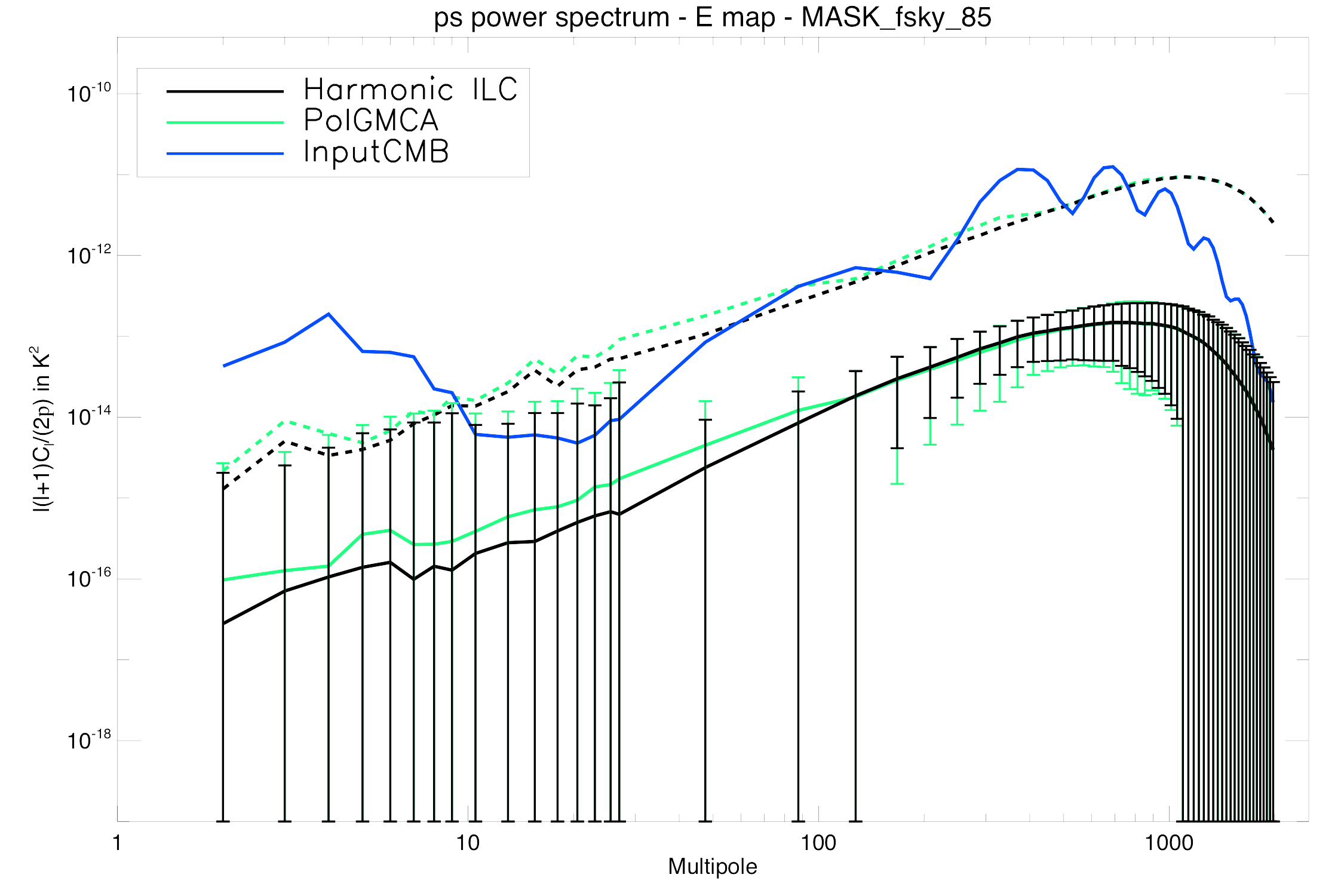}
\includegraphics [scale=0.17]{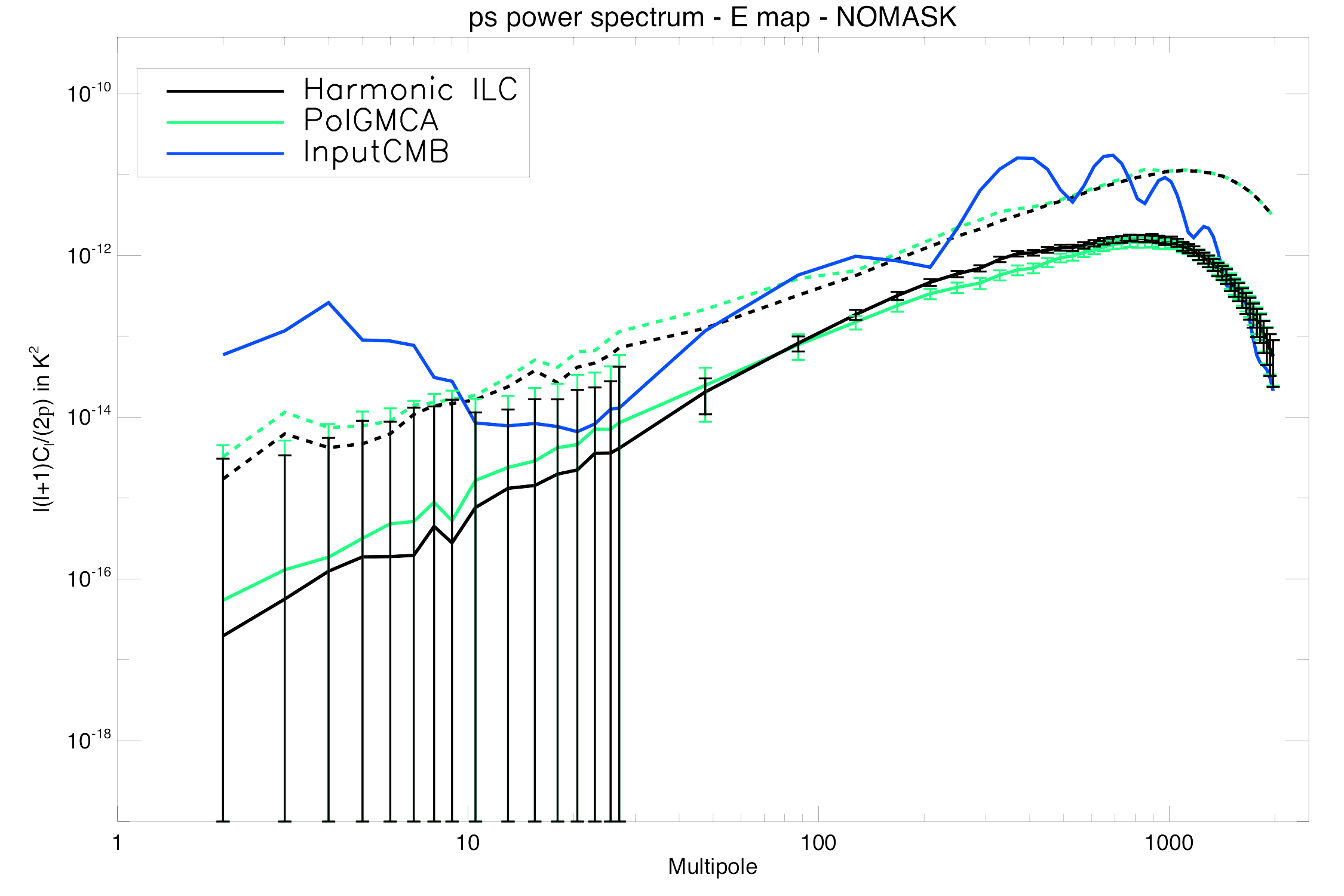}
}}
\caption{{\bf Point sources - top-left :} power spectrum of the point sources residual in the estimated E map from a sky coverage of $25\%$. {\bf top-right :} from a sky coverage of $55\%$. {\bf bottom-left :} from a sky coverage of $85\%$. {\bf bottom-right :} full-sky.}
\label{fig_ps_Q}
\end{figure*}

This paragraph investigates the contamination of the polarized CMB map estimates, and especially the CMB E map from individual foregrounds.\\ 
\begin{itemize}
\item{Thermal dust} Figure~\ref{fig_tdust_Q} features the power spectrum of the residual thermal dust contamination for various sky coverage ranging from $25\%$ to full sky. For intermediate and small scales ({\it i.e.} $\ell > 1000$), thermal dust is by far the major contaminant for both the HILC and PolGMCA E maps, whether it is on the galactic center or at high galactic latitudes. In this range of multipoles, HILC performs slightly better than PolGMCA for sky coverage smaller than $55\%$. For larger sky coverage, the ability of PolGMCA to deliver a  clean estimate of the CMB map on the galactic center greatly helps limiting the increase of thermal dust residual from areas close to the galactic center. It is important to notice that the E map delivered by the PolGMCA algorithm shows a level of thermal dust residual that is about one order of magnitude lower than the level of CMB for $\ell < 1200$, even for a sky coverage as large as $85\%$. \\
At small scales, for $\ell < 100$, the CMB E map computed  with PolGMCA contains significantly lower thermal dust residuals at all latitudes. HILC provides a slightly lower dust level in the range $100 < \ell < 1000$ at high galactic latitudes.\\
\item{Synchrotron} the power spectrum of the synchrotron residual in the HILC and PolGMCA E maps are displayed in Figure~\ref{fig_sync_Q}. HILC a provides lower synchrotron contamination for $\ell < 20$, whether it is on the galactic center or at larger galactic latitudes. However, PolGMCA provides significantly lower synchrotron residuals for $\ell > 100$, even for high galactic latitudes.\\
\item{Point sources} Figure~\ref{fig_ps_Q} shows the power spectrum of the point sources contamination. At large-scale, PolGMCA exhibits a slightly larger amount of point sources residual, for all sky coverage. PolGMCA performs slightly better than HILC at intermediate and small scales. It is interesting to point out that that for large sky coverage, the E map obtained with PolGMCA shows a significantly lower level of point sources residual for $\ell > 100$; this highlights the ability of the PolGMCA algorithm to remove highly non-Gaussian and non-stationary components like point sources, especially on the galactic center. For astrophysical studies, the most prominent point sources will be masked to limit their impact. The level of point sources residual is therefore expected to be significantly lower for both component separation techniques in this setting.
\end{itemize}


\subsection{Towards an accurate full-sky estimate of the CMB polarization maps}

\begin{figure}[htb]
\includegraphics [scale=0.17]{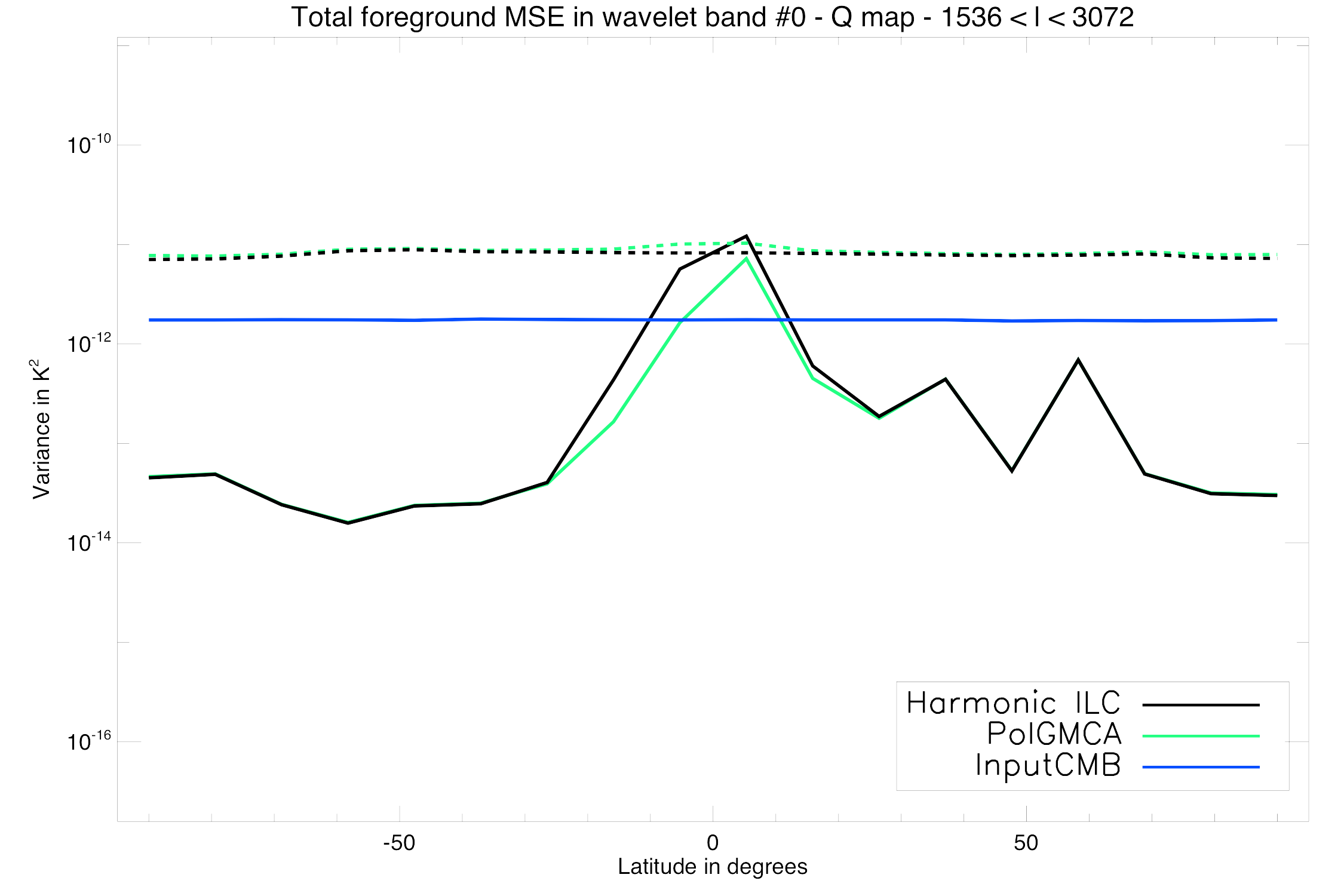}
\hfill
\includegraphics [scale=0.17]{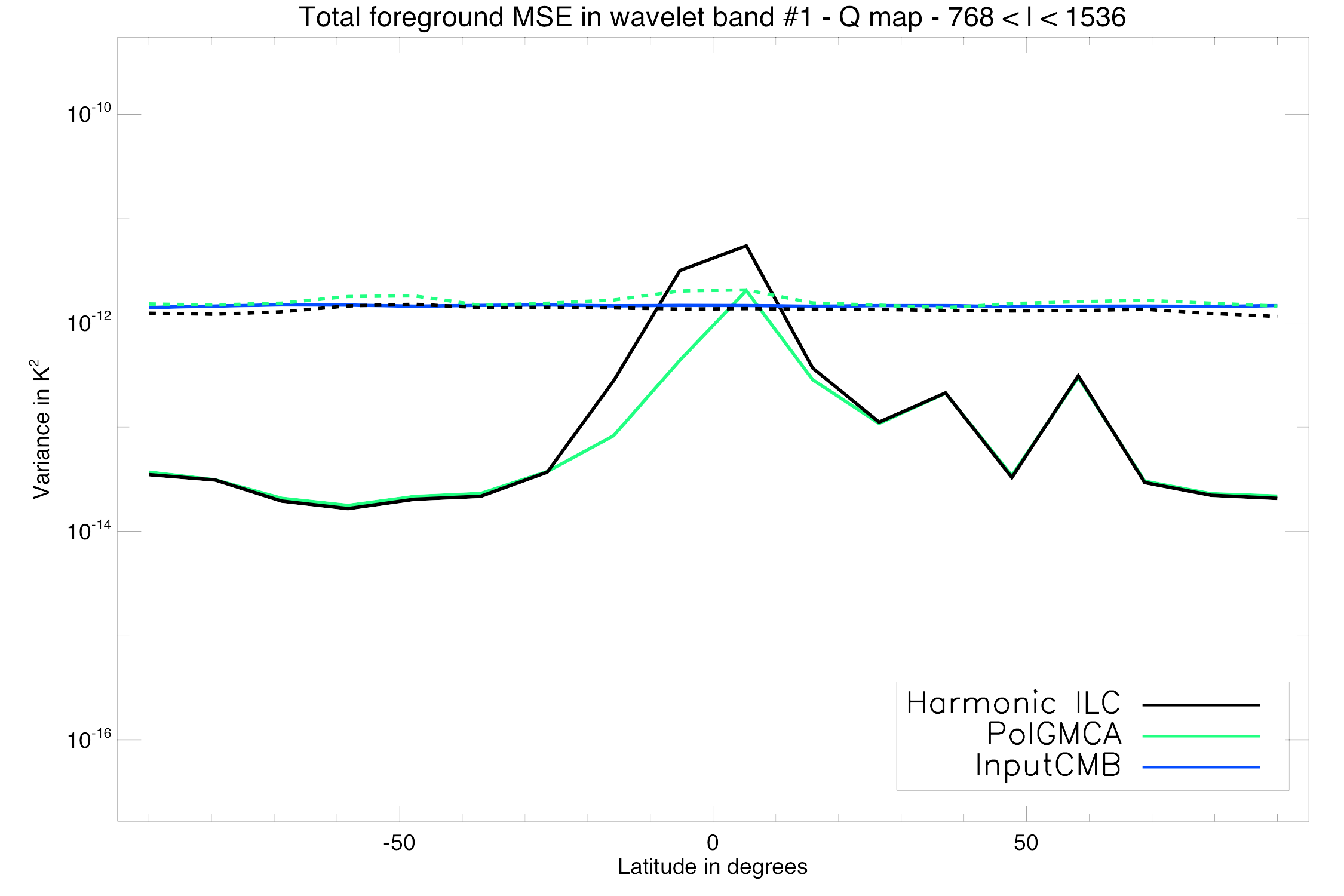}
\includegraphics [scale=0.17]{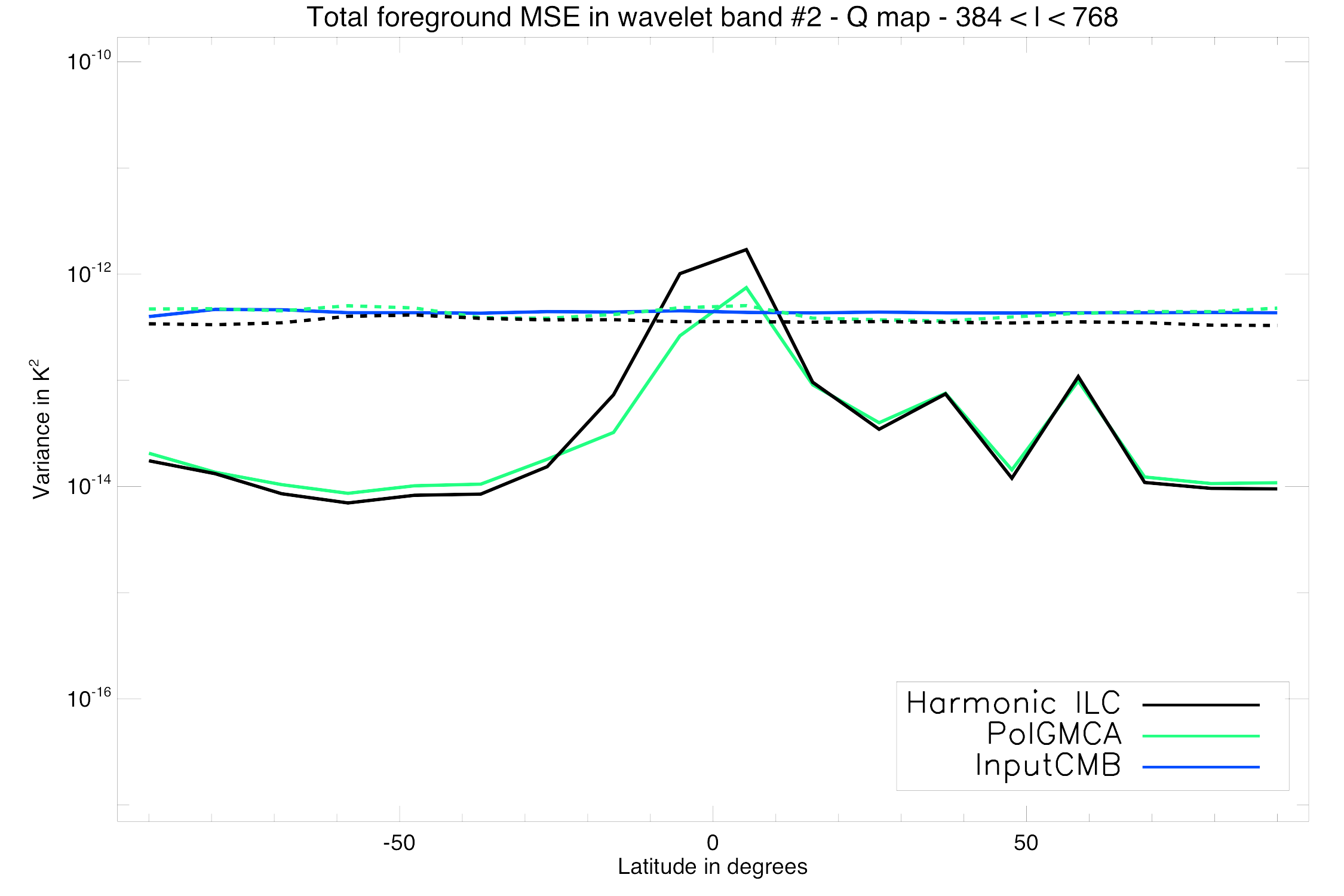}
\hfill
\includegraphics [scale=0.17]{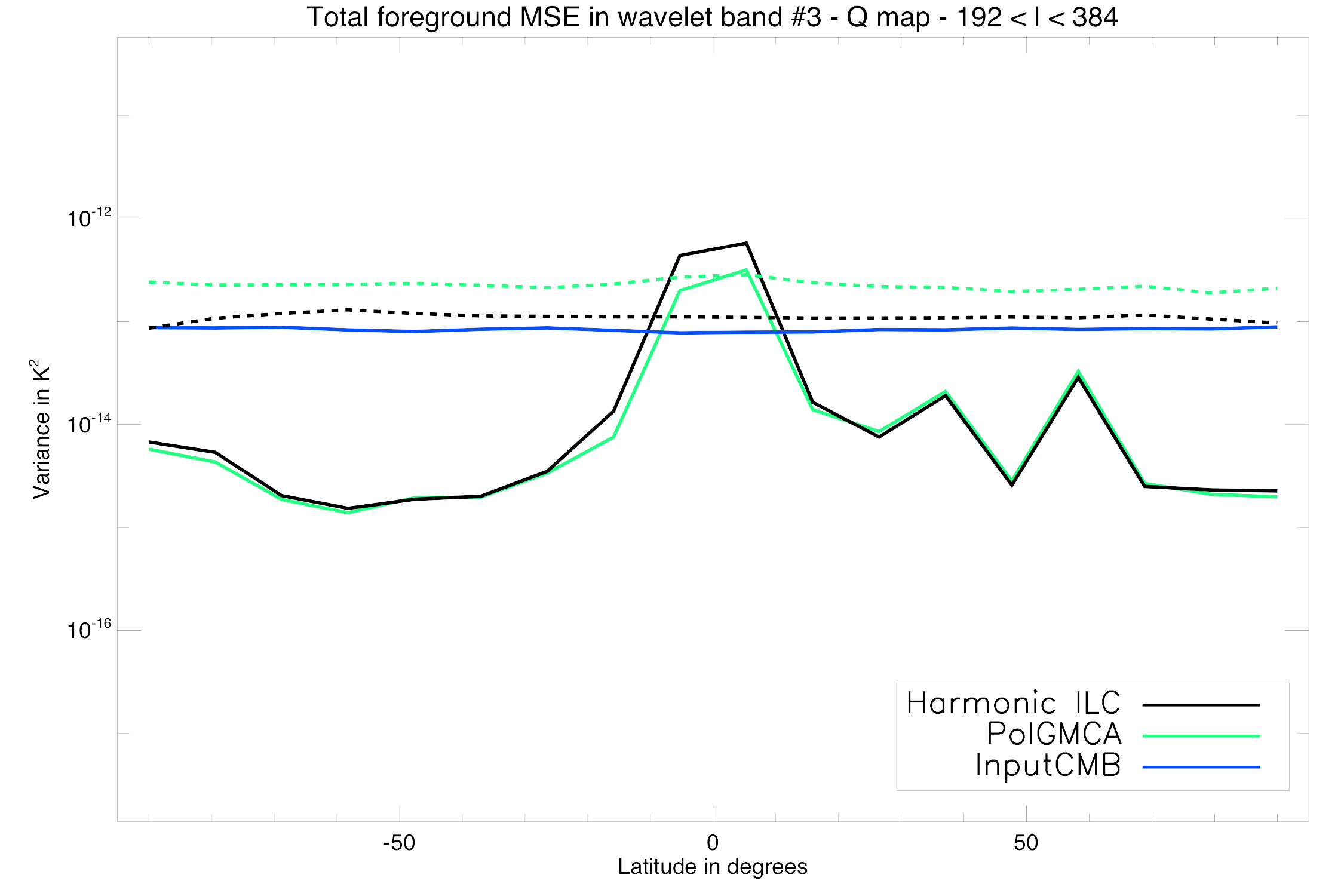}
\includegraphics [scale=0.17]{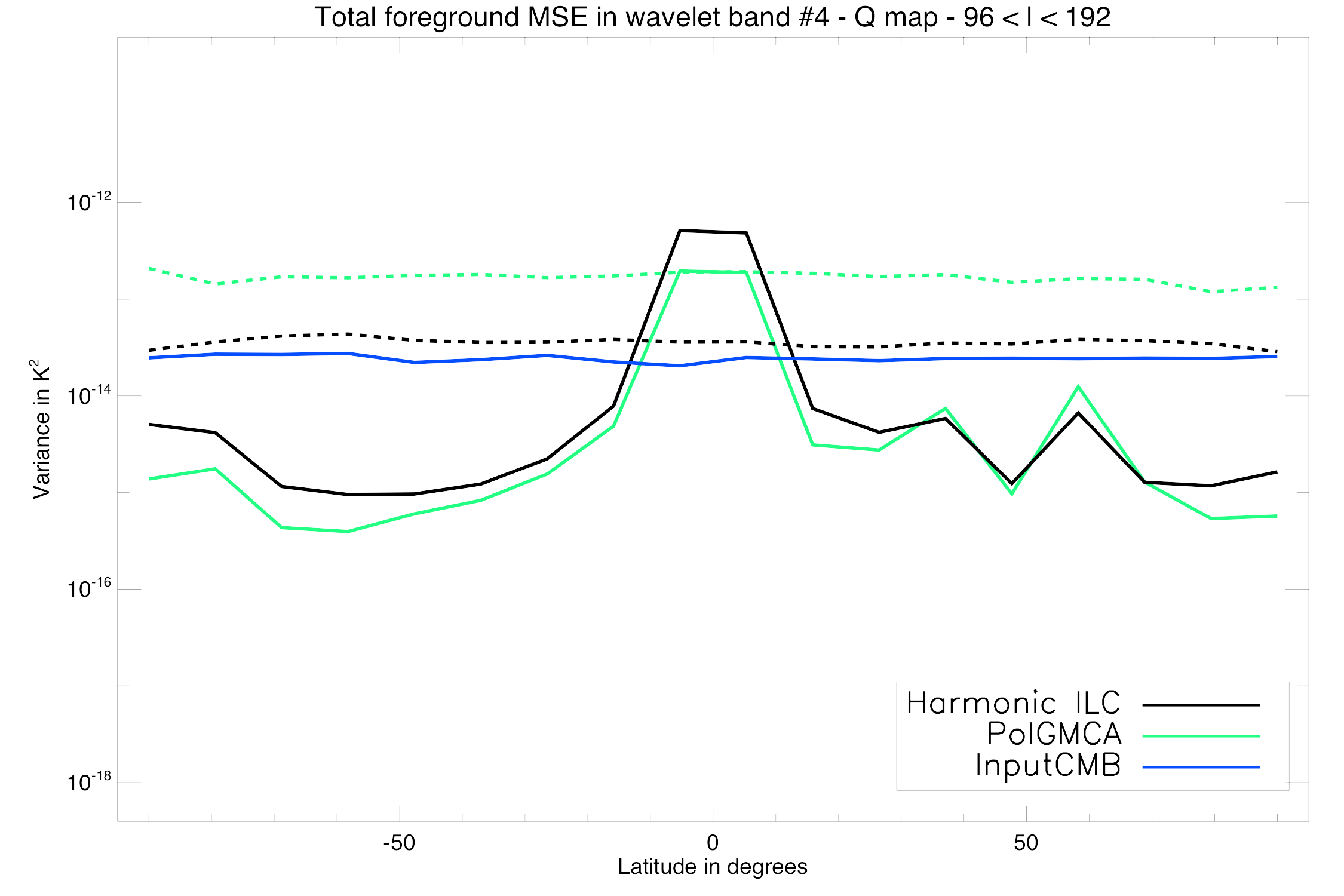}
\hfill
\includegraphics [scale=0.17]{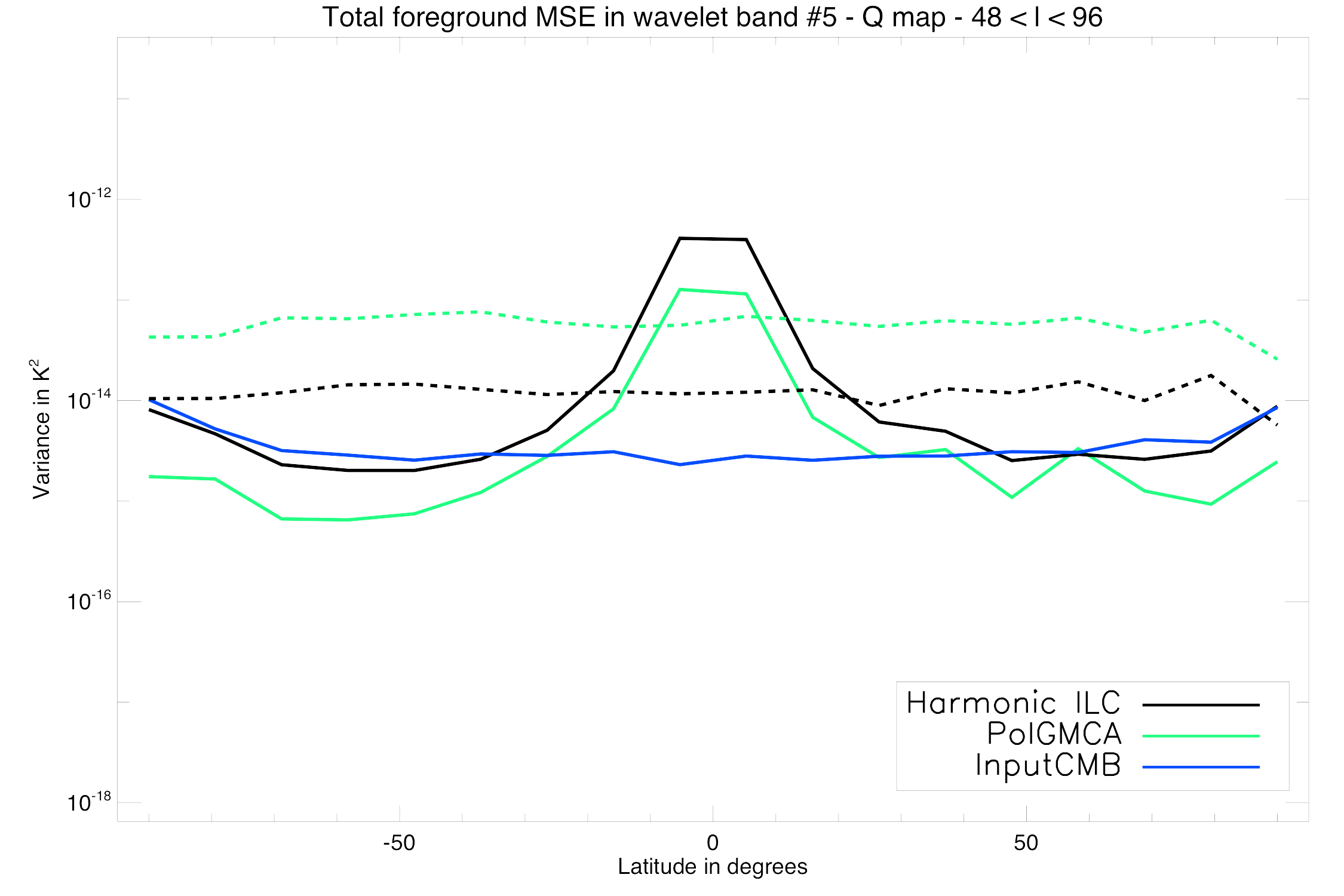}
\caption{{\bf Total foregrounds residuals in the Stokes parameter Q map: } mean squared error (MSE) per band of latitude in the wavelet domain. Dashed lines represent the noise level for each component separation method.}
\label{fig_total_frg_mse_Q}
\end{figure}

In this section, we investigate with more precision the quality of estimation of the CMB polarization map across the sky. Indeed, one of the objectives of this article is to allow the accurate full-sky estimation of polarized CMB maps. For that purpose, and following an analysis method we have used \citep{2012arXiv1206.1773B,WMAP9_LGMCA, PR1_LGMCA}, the accuracy of CMB map can be evaluated by computing the level of foreground residuals in bands of latitudes, in the wavelet domain. Using wavelets, one can analyze the estimated CMB maps with a good localization in space as well as in harmonic space.\\
More precisely, the estimated Q and U maps have been decomposed in $6$ wavelet bands -- see \citep{Starck05}. Each of the $6$ bands is then split uniformly into $18$ bands of latitudes; each band of latitude has width $10^\circ$. Figure~\ref{fig_total_frg_mse_Q} reports the level of energy (Euclidean norm) of the total foreground residuals ({\it i.e.} thermal dust, synchrotron, and point sources) for the $6$ wavelet scales. By convention, the wavelet scale $0$ corresponds to the smallest scales ($\ell >  1600$) and scale $5$ to the large scales ($50 < \ell < 100 $).\\
It is first interesting to notice that, at all scales, HILC has a level of foreground residuals that exceed the level of CMB on the galactic center, in latitudes of about $\pm 15^\circ$. This is precisely at the vicinity of the galactic center that the stationarity assumptions, made in the framework of HILC, is probably the least valid and where the use of spatial localization and sparsity is the most beneficial for component separation. Consequently, the CMB map estimated by the PolGMCA algorithm has significantly lower foreground residuals for galactic latitudes lower than $15^\circ$. More precisely, in the galactic plane, HILC Q map has an order of magnitude higher foreground residuals than PolGMCA, in wavelet bands corresponding to $\ell > 400$.\\
For the first three wavelet scales, the PolGMCA Q and U maps have a level of foreground residuals which is constantly lower than the level of CMB all over the sky. At larger scales, CMB is dominated by the foregrounds only for latitudes which are lower than $\pm 10^\circ$.\\
For galactic latitudes larger than $\pm 25^\circ$, these plots confirm the very good efficiency of the HILC algorithm in regions where instrumental noise is, by far, the most prominent contaminant in the data. The PolGMCA algorithm generally provide similar or slightly higher levels of foregrounds in high galactic regions, but still much lower than the level of instrumental noise.\\
These numerical results highlight the efficiency of the proposed PolGMCA algorithm to provide a very accurate estimation of the polarized CMB maps at the vicinity of the galactic center while providing residual levels that are competitive to the very effective HILC algorithm at high galactic latitudes at intermediate and small scales. The PolGMCA is shown to be particularly effective at large-scales ({\it i.e.} for $\ell < 200$), where it yields estimates of the CMB Q and U maps with significantly lower levels of total foreground at all galactic latitudes.\\ \\

\begin{figure}[htb]
\includegraphics[scale=0.17]{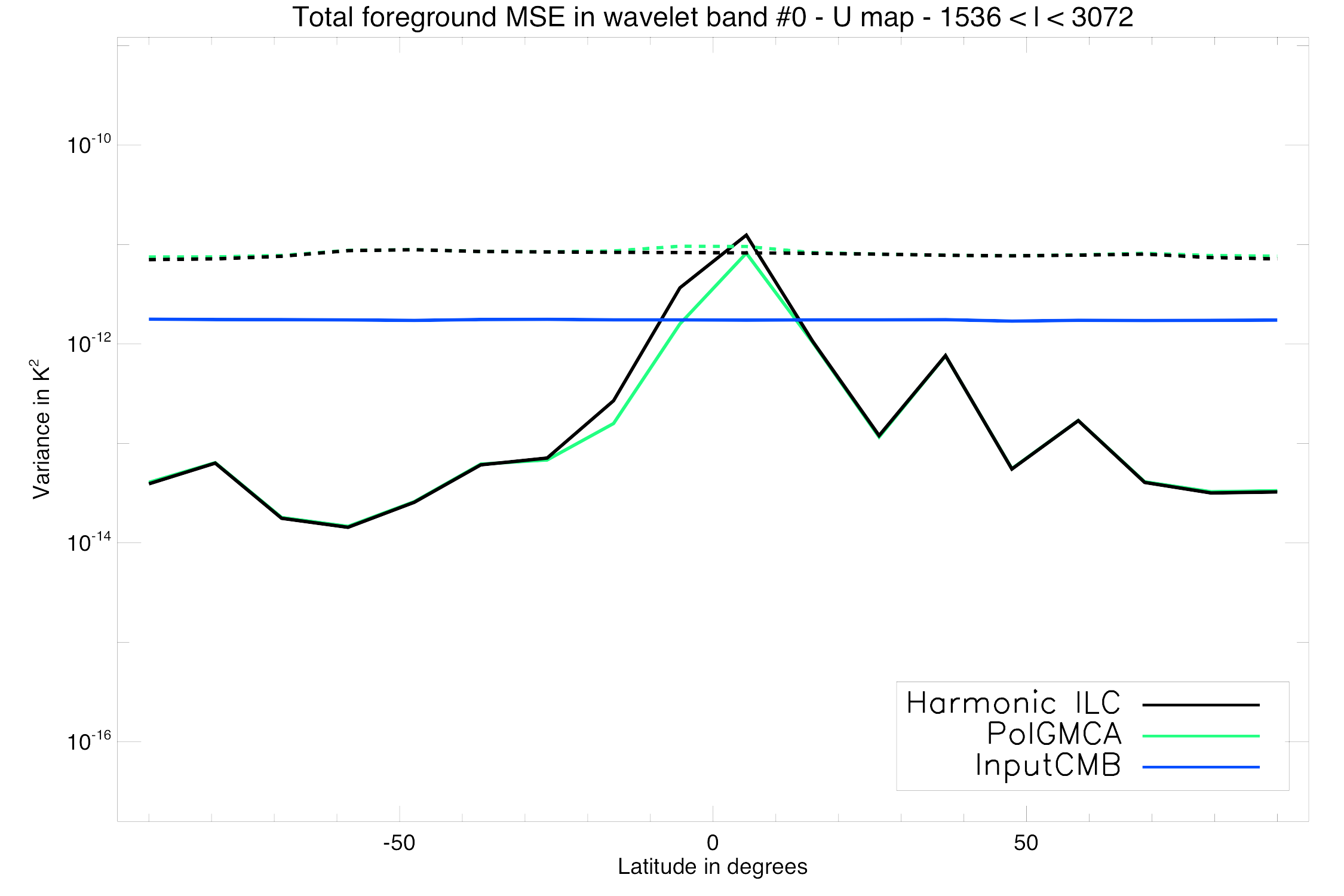}
\hfill
\includegraphics[scale=0.17]{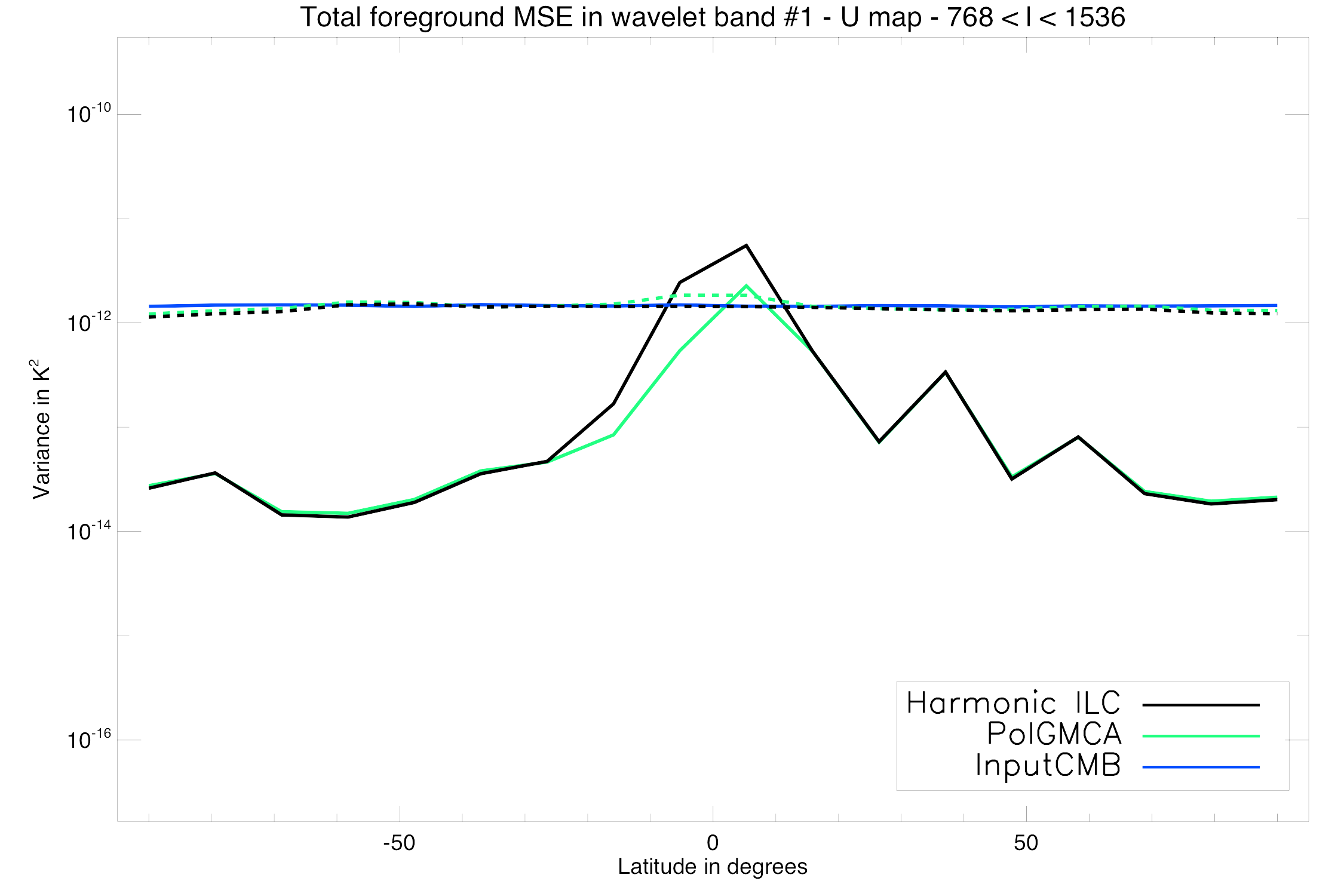}
\includegraphics[scale=0.17]{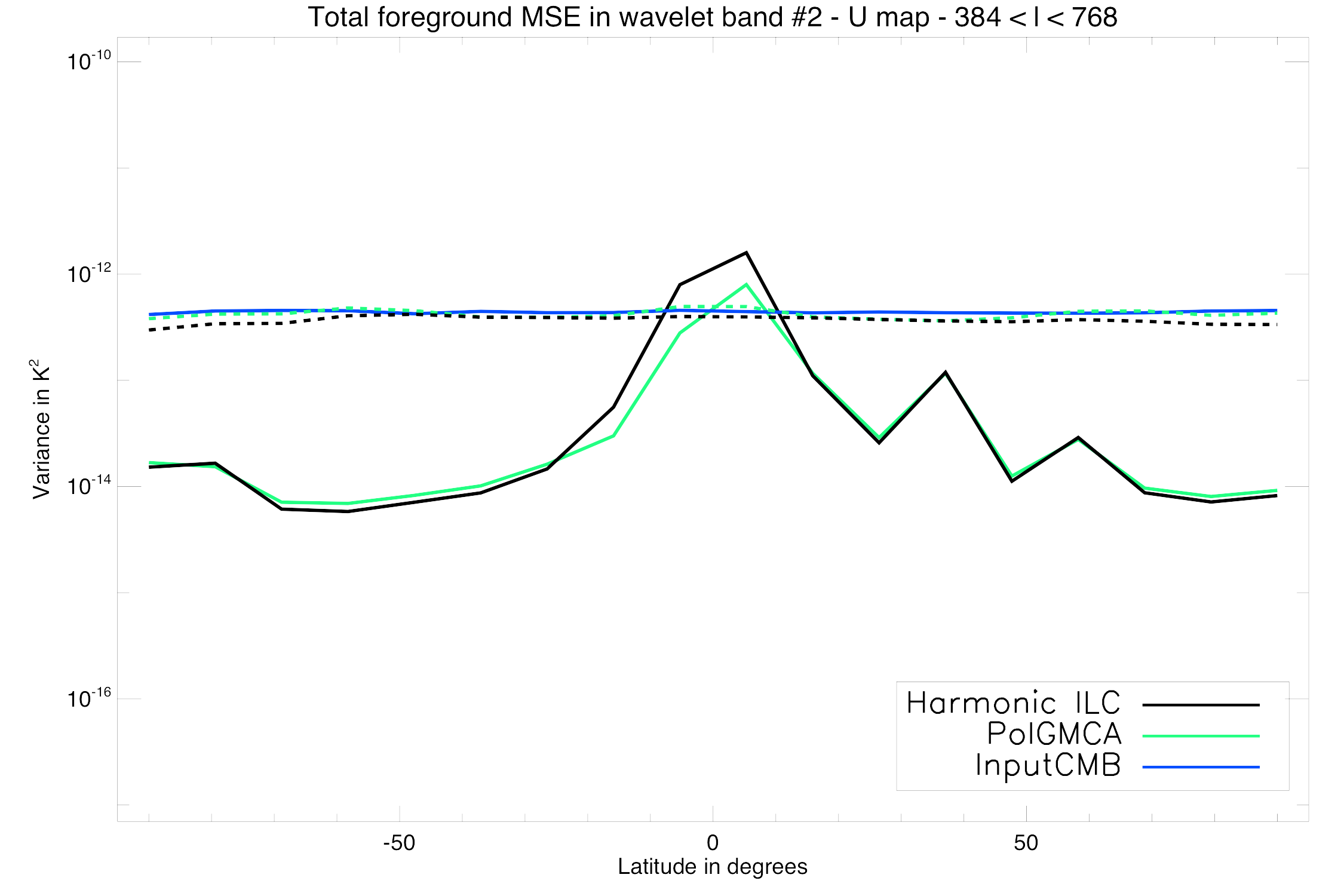}
\hfill
\includegraphics[scale=0.17]{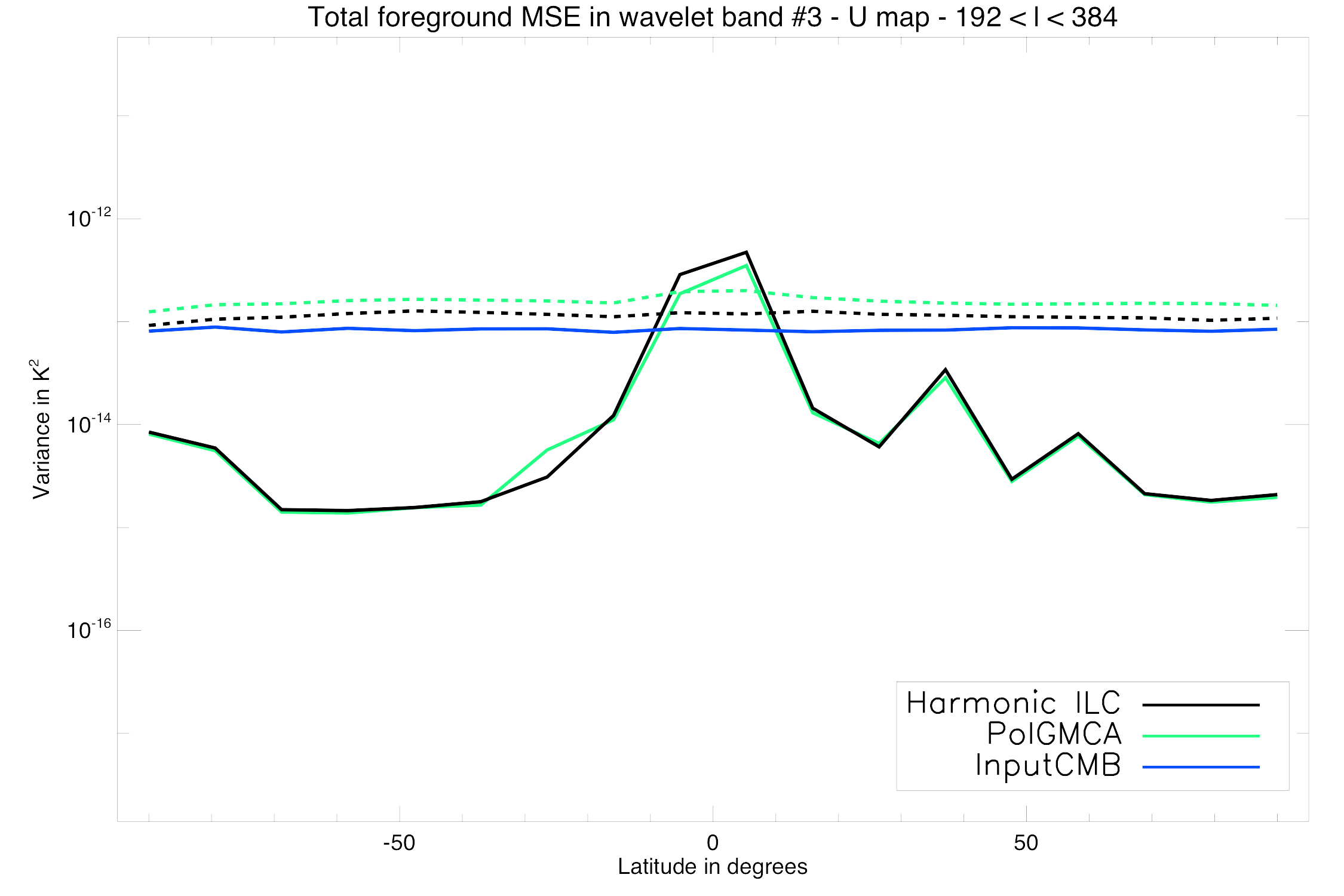}
\includegraphics[scale=0.17]{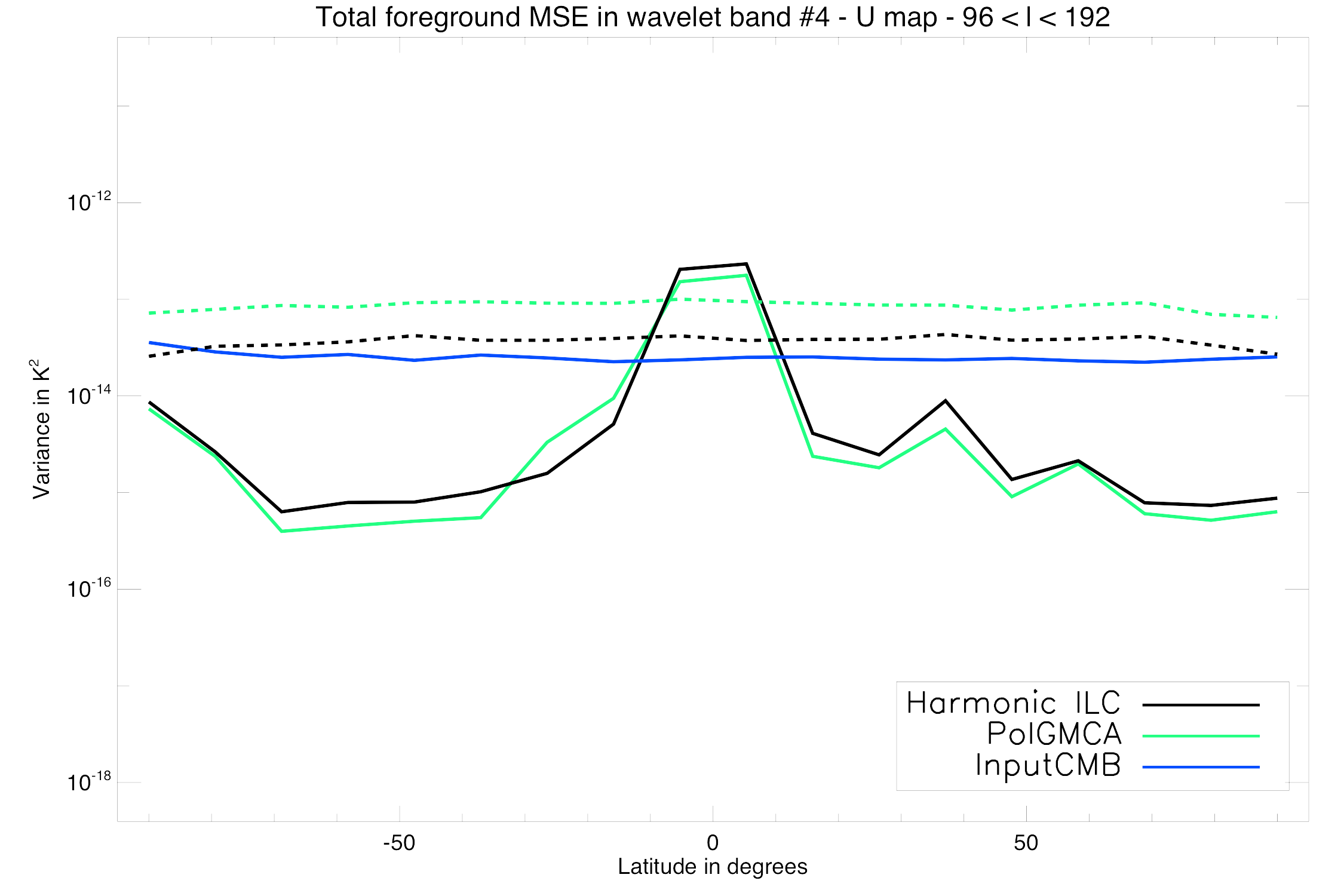}
\hfill
\includegraphics[scale=0.17]{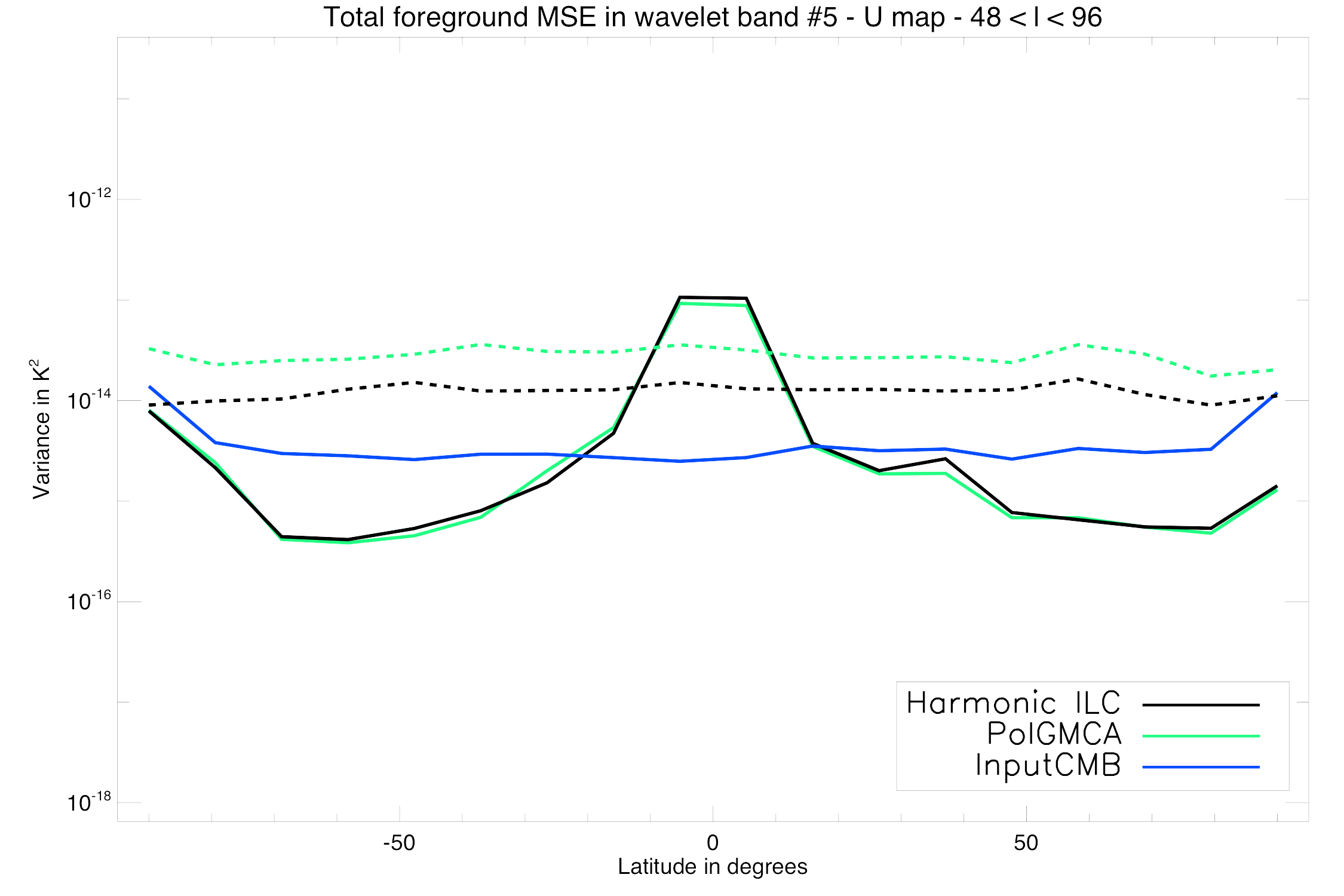}
\caption{{\bf Total foregrounds residuals in the Stokes parameter U map: } mean squared error (MSE) per band of latitude in the wavelet domain. Dashed lines represent the noise level for each component separation method.}
\label{fig_total_frg_mse_U}
\end{figure}



\section{Conclusion}
We present a novel component separation method which is dedicated to the estimation of the CMB polarization modes. This approach builds upon the GMCA algorithm, which emphasizes on the sparsity of the sought after sources in the wavelet domain to separate them out. The proposed dubbed PolGMCA extends the basic GMCA algorithm with two significant improvements: i) it allows for the efficient separation of partially correlated sources thanks to recent improvements of the GMCA algorithm, and  ii) it aggregates sparse and second-order based CMB estimators to benefit from their complementary characteristics in terms of propagated noise and foreground residuals. Subsequently, we show that the proposed PolGMCA algorithm exhibit the following interesting properties :
\begin{itemize}
\item \cjb{It provides an improved estimation of the polarized CMB maps particularly at large-scales, typically for $\ell < 200$, outside the galactic center, where the ability to account for the partial correlation the components is the essential. However, these improvements are not significant with respect to the noise-related uncertainty.}  
\item It provides very accurate estimates of the polarized CMB maps in regions where foregrounds, which are by nature non-Gaussian and non-stationary, dominate. This is specifically the case at the vicinity of the galactic center at latitudes $\pm 15^\circ$. In this setting, the PolGMCA algorithm benefits to a large amount of its ability to account for the partial correlation of the components to be removed
\item In noise-dominated regions of the sky, the polarized CMB maps provided by the PolGMCA algorithm have levels of foregrounds which are close to the HILC algorithm, which is assumed to be close to optimality in this noise regime.
\end{itemize}
Altogether, these results make the PolGMCA algorithm a good candidate for galactic studies in polarization, studying large-scale anomalies and testing the detection of B-modes from Planck data.

\begin{acknowledgements} 
This work is supported by the European Research Council grant SparseAstro (ERC-228261) and partially by the French National Agency for Research (ANR) 11-ASTR-034-02-MultID.
 We used Healpix software \citep{Healpix},
  iSAP\footnote{\url{http://jstarck.free.fr/isap.html}} software. 
\end{acknowledgements}

\bibliographystyle{aa} 
\bibliography{gmca_bib}

\vspace{-.5cm}
\appendix

%
%
%
%

%
\begin{appendix}

\section{Harmonic ILC implementation}
\label{sec:HILC} 
\cjb{Harmonic ILC has been widely used, mainly to estimate CMB maps from WMAP data in temperature data \citep{WMAP_Tegmark_03} and polarization \citep{WMAP_Naselsky_Pola}. More recently, the SMICA CMB polarized maps derived from the Planck PR2 data \citep{PR2_compsep} are equivalent to a HILC solution at high $\ell$ multipoles. The appeal of HILC resides in its ability to precisely account for the different beams at each channel as well as its accuracy to model stationary Gaussian random fields like the CMB.\\
In this paper, and inspired by \citep{WMAP_Tegmark_03,WMAP_Naselsky_Pola}, we have implemented a HILC method that performs into two steps:}
\begin{itemize}
\item{\it Fitting the ILC weights:} \cjb{ILC estimates weights to obtain a minimum variance solution. In the case of HILC, this step is performed independently on overlapping bands of multipoles that span the desired range of multipoles $[0,2500]$. The bands have a fixed width $\Delta = 80$; this width has been tuned so as to provide the best estimation accuracy in PSM simulations, as a trade-off between good foreground modeling (narrow bands) and statistical robustness (broad bands). All the $81$ bands used to derive ILC weights are displayed on the left panel of Figure~\ref{fig:hilc_bands}. Let $b_{i,\ell}$ be the beam of the $i$-th frequency channel, $w_{l,\ell}$ be the filter defining the $l$-th band, and $\tilde{x}_{i,(\ell,m)}$ be the $a_{\ell,m}$ coefficient of the $i$-th frequency channel, the weights $[\pi_{i,l}]_{i=1,\cdots,M}$ in the $l$-th band are computed by minimizing the following problem: }
\begin{equation}
\min_{\{\pi_{i,l}\}_{i=1,\cdots,M}} \sum_{\ell,m}\left \| \sum_i w_{l,\ell}\pi_{i,l} b_{i,\ell} \tilde{x}_{i,\ell,m} \right \|_{\ell_2}^2  \quad \mbox{ s.t. } \quad \sum_{i=1}^M \pi_{i,l} = 1,
\end{equation}

\begin{figure}[htb]
\includegraphics [scale=0.17]{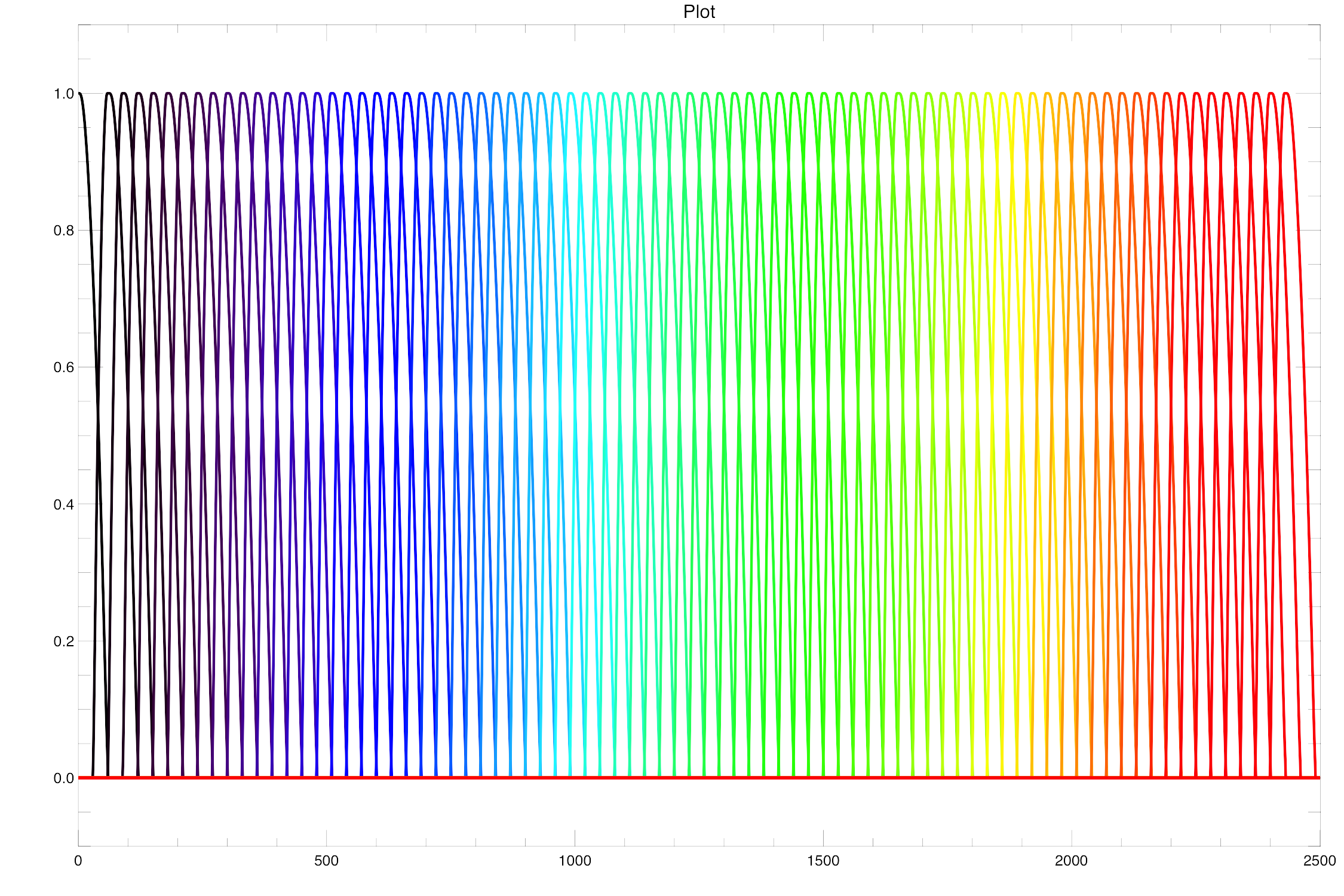}
\includegraphics [scale=0.17]{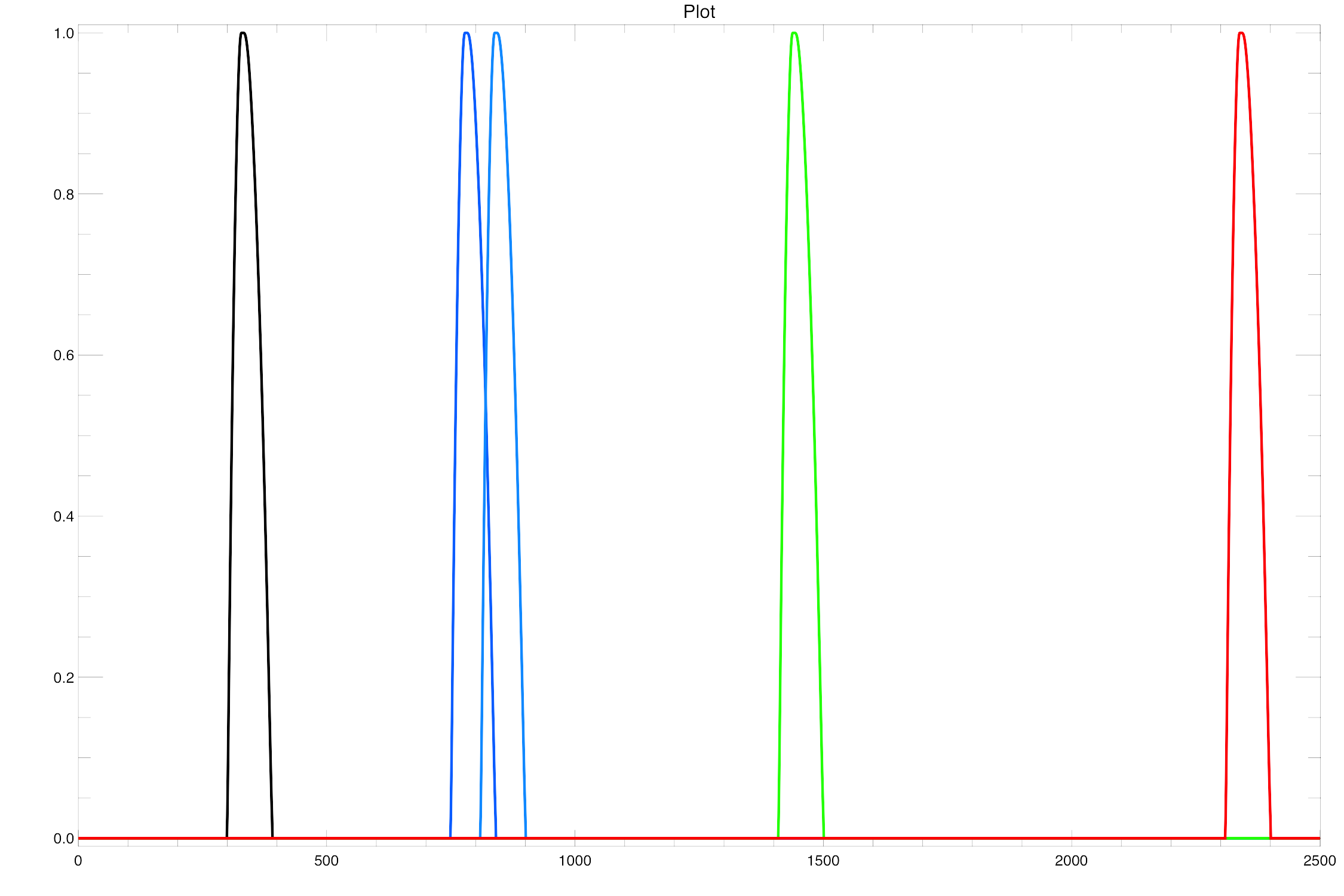}
\caption{{\bf Bands of multipoles used to derive HILC weights: left}  All the $81$ overlapping bands used to compute the ILC weights in the harmonic domain. {Right: } examples of a $5$ different bands for better visualization.}
\label{fig:hilc_bands}
\end{figure}
\item{\it Deriving $\ell$-dependent weights:} \cjb{ from the $81$ bands, one derives weight per bands and observations. A simple interpolation procedure is used so as to derive a weight per $\ell$ and per observation $i$. This interpolation step is performed to provide weight vectors $\pi_{i} = [\pi_i(0), \cdots,\pi_i(\ell), \cdots, \pi_i(\ell_{\max})]$ as follows:
\begin{equation}
\forall i=1,\cdots,M; \quad \pi_i = \frac{\sum_l w_{l,\ell} \pi_{i,\ell}}{\sum_l w_{l,\ell}}
\end{equation}
which is essentially a simple interpolation by weighted averaging, where the filters $w_{l}$ are used as weights.}
\end{itemize}
\cjb{The exact same HILC algorithm is used to aggregate the CMB estimators in the PolGMCA algorithm described in \ref{sec:compsep}.}
\end{appendix}

\end{document}